\documentclass[10pt,twocolumn,superscriptaddress,english,10pt,prl,showpacs,floatfix,aps]{revtex4-1}
\usepackage[latin9]{inputenc}
\usepackage{amsthm}
\usepackage{amsmath}
\usepackage{amssymb}
\usepackage{esint}
\usepackage{graphicx}
\usepackage{upgreek}
\usepackage{xspace}

\makeatletter

\@ifundefined{textcolor}{}
{%
 \definecolor{BLACK}{gray}{0}
 \definecolor{WHITE}{gray}{1}
 \definecolor{RED}{rgb}{1,0,0}
 \definecolor{GREEN}{rgb}{0,1,0}
 \definecolor{BLUE}{rgb}{0,0,1}
 \definecolor{CYAN}{cmyk}{1,0,0,0}
 \definecolor{MAGENTA}{cmyk}{0,1,0,0}
 \definecolor{YELLOW}{cmyk}{0,0,1,0}
}

\usepackage{soul}
\usepackage{braket}
\usepackage{hyperref}

\renewcommand{\v}[1]{\ensuremath{\mathbf{#1}}} 
 
\let\baraccent=\= 
\renewcommand{\=}[1]{\stackrel{#1}{=}} 

\newcommand{\didv}{d$I/$d$V$\xspace}

\newcommand{\DeltaT}{\ensuremath{\Delta_\mathrm{tip}}\xspace}
\newcommand{\DeltaS}{\ensuremath{\Delta_\mathrm{sample}}\xspace}
\newcommand{\Fig}[1]{Fig.~\ref{fig:#1}}
\newcommand{\Figure}[1]{Figure~\ref{fig:#1}}

\newcommand{\YSR}{YSR~}

\DeclareMathOperator{\unitspace}{\xspace}

\DeclareMathOperator{\umueV}{\unitspace\mathrm{\mu eV}}

\DeclareMathOperator{\uK}{\unitspace\mathrm{K}}

\makeatother

\usepackage{babel}

\newcommand{\eg}{{\it e.g.}}
\newcommand{\ie}{{\it i.e.}}
 \newcommand{\vect}[1]{\boldsymbol{#1}}

\begin{document}

\author{Benjamin W. Heinrich}\email{bheinrich@physik.fu-berlin.de}
\affiliation{Fachbereich Physik, Freie Universit\"at Berlin, 14195 Berlin, Germany}

\author{Jose I.\ Pascual}\email{ji.pascual@nanogune.eu}
\affiliation{CIC nanoGUNE, 20018 Donostia-San Sebasti\'an, Spain} 
\affiliation{Ikerbasque, Basque Foundation for Science, 48011 Bilbao, Spain}

\author{Katharina J. Franke}\email{franke@physik.fu-berlin.de}
\affiliation{Fachbereich Physik, Freie Universit\"at Berlin, 14195 Berlin, Germany}
\date{\today}

\title{Single magnetic adsorbates on $s$-wave superconductors}

\begin{abstract}
In superconductors, magnetic impurities induce a pair-breaking potential for Cooper pairs, which locally affects the Bogoliubov quasiparticles and gives rise to Yu-Shiba-Rusinov (\YSR or Shiba, in short) bound states in the density of states (DoS). These states carry information on the magnetic coupling strength of the impurity with the superconductor, which determines the many-body ground state properties of the system. Recently, the interest in Shiba physics was boosted by the prediction of topological superconductivity and Majorana modes in magnetically coupled chains and arrays of Shiba impurities.

Here, we review the physical insights obtained by scanning tunneling microscopy into single magnetic adsorbates on the $s$-wave superconductor lead (Pb). We explore the tunneling processes into Shiba states, show how magnetic anisotropy affects many-body excitations, and determine the crossing of the many-body groundstate through a quantum phase transition. Finally, we discuss the coupling of impurities into dimers and chains and their relation to Majorana physics.

\end{abstract}

\maketitle

\tableofcontents 

\section{Introduction}

Over the last decade, the interest of condensed matter physicists in low-dimensional hybrid superconductor-magnet systems increased dramatically. It was realized that these are prime candidates to bear topological phases and, in particular, Majorana zero modes.  These quasiparticles are envisioned as building blocks for the construction of topologically protected qubits, which could make quantum computing more fault-tolerant~\cite{Alicea2012,Beenakker2013,Elliott14}. Such topological phases can be constructed on the basis of magnetic atoms or molecules at the interface to a superconductor. These magnetic {\it impurities} exhibit a scattering potential to the quasiparticles of the substrate, which leads to bound states, the so-called Yu-Shiba-Rusinov (\YSR or Shiba) states~\cite{Yu1965,Shiba1968,Rusinov1968,Rusinov1969}. While their theoretical investigation dates back to as far as the 1960s, their experimental characterization remained in its infancy until very recently. The renewed interest stems, on the one hand, from the arising of new fundamental aspects in connection with topological superconductors   and, on the other,  from advances of recent technologies to resolve their intriguing properties. These developments enable the {\it in vacuo} preparation of clean superconductor surfaces, the deposition of single atoms at low temperatures, and their investigation with high spatial and energy resolution at temperatures well below the critical temperature ($T_{\text{c}}$) of the superconductor.    
After the pioneering work from the Don Eigler group in 1997~\cite{Yazdani1997} on single Manganese (Mn) and Gadolinium (Gd) adatoms on a Niobium (Nb) single crystal surface at $4$\,K, it took more than ten years until new experiments with increased energy resolution were published~\cite{Ji2008,Ji2010a}. Experiments at lower temperatures and the use of superconducting tips enabled a more detailed analysis of \YSR bound states induced by single paramagnetic adsorbates on the $s$-wave superconductor Pb. 
In addition, the proposal of building topologically protected states from magnetic adsorbates on superconductors has pushed the field even further. 
These states can be designed by coupling \YSR states along one-dimensional chains. For understanding the formation of \YSR bands and proximity-induced p-wave superconductivity within these chains, a basic understanding of the individual building blocks and their interaction is required.

The first theoretical descriptions started with classical spin models and already predicted the spectral properties and distance dependence of the \YSR wave function~\cite{Yu1965,Shiba1968,Rusinov1968,Rusinov1969,Flatte1997PRL}. Models of quantum impurities, \textit{i.e.}, impurities with internal degrees of freedom that can be activated by spin-flip scattering of the host material, require much more demanding calculations. In practice, approximate methods such as mean field calculations~\cite{Salkola1997,Rodero2012}, perturbation theory~\cite{Zonda2015,Zonda2016}, and numerical renormalization group theory~\cite{Shiba1993, Bauer2007,Zitko2017} are employed. These give valuable insights into the spectral properties at and around the impurity site. The complexity of crystal-field splitting~\cite{Moca2008, Zitko2011}, vibrational degrees of freedom~\cite{Golez2012}, spin-orbit coupling~\cite{Kim2015,Pershoguba2015,Pershoguba2016,Kaladzhyan2016}, temperature~\cite{Zitko2016a}, or external magnetic fields~\cite{Zitko2011} have also been included into the theoretical treatments of single impurities. Theoretical models also discussed the coupling of two magnetic impurities long before their experimental observation~\cite{Flatte2000,Morr2003,Morr2003a,Morr2006,Yao2014,Yao2014a,Zitko2015,Meng2015}. Finally, the theoretical considerations expanded on the coupling of magnetic impurities into one-dimensional chains and led to the prediction of Majorana bound states for helical~\cite{Nadj-Perge2013,Pientka2013} and ferromagnetic chains~\cite{Nadj-Perge2014, Li2014}. Very recently, exotic states have also been predicted in two-dimensional lattices of \YSR impurities~\cite{Nakosai2013,Rontynen2015,Li20162D,Menard2016,Rachel2017,Schecter2017}.

Hence, theory anticipates many interesting phenomena, which have become verifiable only with the technological advances in scanning probe techniques over the last few years. In this overview article, we review the progress made in this field from an experimental perspective. 

In Section~\ref{intro}, we will start with a short introduction to the physical concepts of \YSR states but refrain from a detailed theoretical treatment, which can be found, \eg, in the review by Balatsky and coworkers~\cite{Balatsky2006}. We then will discuss the experimental detection mechanism of \YSR states in tunneling experiments, their relation to Andreev bound states as observed in superconducting quantum dot experiments, and  the transport processes through subgap states in Sec.~\ref{Detection}. 
In Section~\ref{Sec:orbital} we review the orbital nature of these states. In Sec.~\ref{Sec:QP}, we explore the effect of magnetocrystalline anisotropy on the many-body phase diagram and determine the ground state's spin, which undergoes a quantum phase transition caused by a varying exchange coupling strength. In Sec.~\ref{Sec:extension} we detail the origin of extended wave functions of the bound states. 
Finally, in Sec.~\ref{Sec:coupling}, we explore the hybridization of single impurities and will end the review with a brief outlook on future research directions in Sec.~\ref{sum}.

\section{Yu-Shiba-Rusinov bound states in $S$-wave superconductors}
\label{intro}

Yu-Shiba-Rusinov bound states are a result of an exchange scattering potential, as we describe in the following in more detail. Here, we shall start with a note on the terminology: Andreev bound states describe a general scattering potential, \ie, not necessarily of magnetic origin. 
In the case of exchange scattering, the quantum dot community typically also refers to Andreev bound states for historical reasons, whereas the magnetic-adatom community names these bound states as \YSR or Shiba states.

The presence of paramagnetic impurities in a superconductor is detrimental for superconductivity. Already in the 1950's Matthias found that the addition of rare earth atoms to the bulk of a superconductor reduces the critical temperature T$_c$ proportionally to the concentration of impurities \cite{Matthias1958}. 
Abrikosov and Gor´kov developed a microscopic theory for high impurity concentrations  considering the effect of a scattering potential with broken time-reversal symmetry~\cite{Abrikosov60}. A localized spin adds an exchange component to the scattering with Cooper pairs, which tends to misalign their spins and, thus, behaves as a pair-breaking potential. 

The exchange scattering of a spin with Cooper pairs is described by  a term $JS_{imp}$, with $J$ being the exchange coupling strength of the spin state $S_{imp}$ of the impurity with conducting electrons of the substrate. This term adds up to the (non-magnetic) component  of the scattering, described by the (time-reversal invariant) potential term $U$. 

In the limit of large impurity densities, the local spins  yield a gradual reduction of the superconducting gap with concentration. The gap then closes before the superconducting state is fully suppressed. However, in the  dilute limit of  impurity concentrations with an average separation larger than the superconducting coherence length, the scattering amplitudes $J$ and $U$ compete with the superconducting pairing, which results in bound states localized at the impurity sites.

\subsection{Dilute limit of paramagnetic impurities on a superconductor}
\label{sec:theory}

\begin{figure}[t]
\includegraphics[width=1\columnwidth]{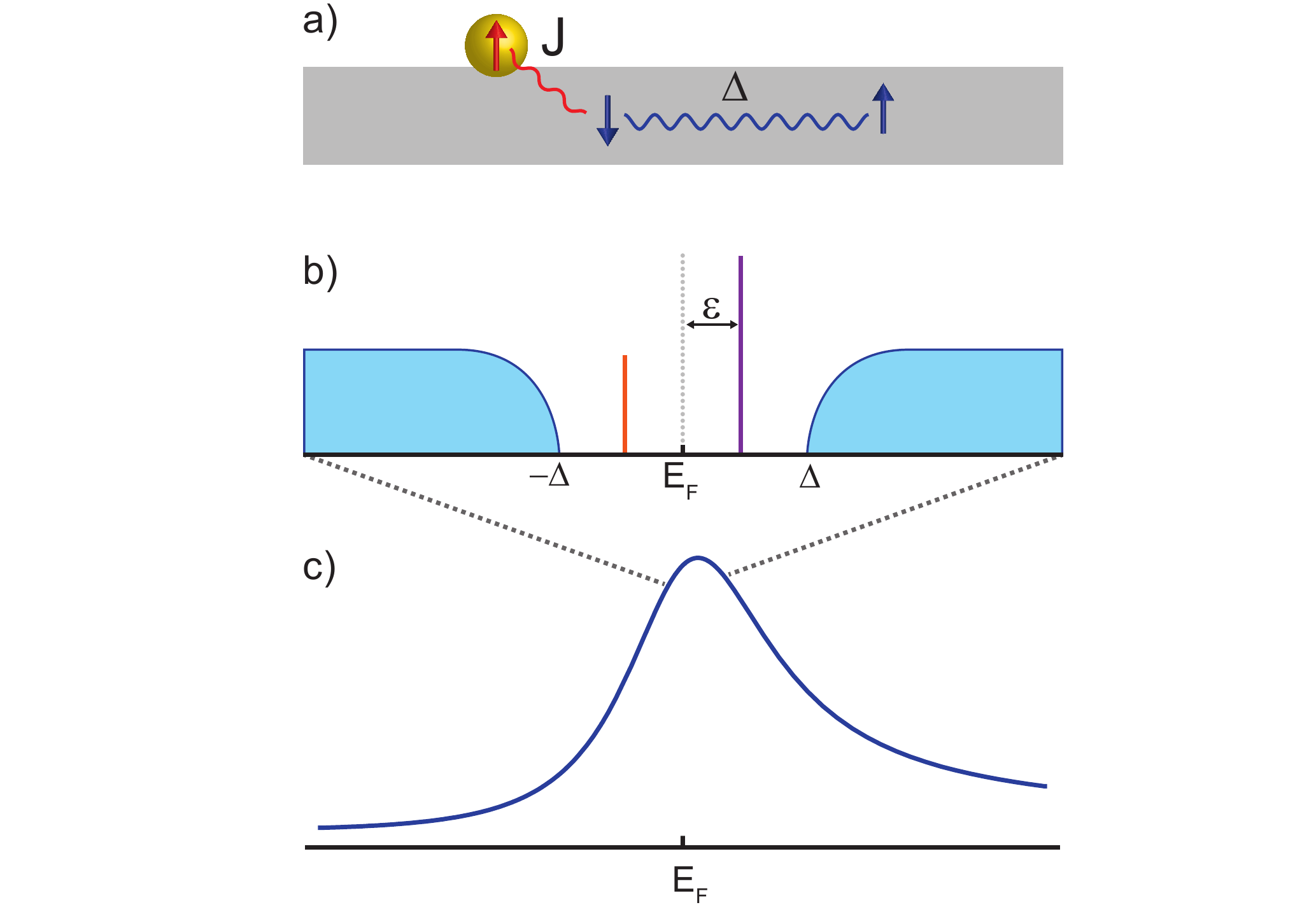}
\caption{a) A paramagnetic impurity in contact with an $s$-wave superconductor. The electron-phonon coupling provides a positive interaction $\Delta$ between electrons of opposite spin and momentum and couples these into Cooper pairs. The impurity spin is exchange coupled with strength $J$ to the Cooper pairs. It locally breaks time-reversal symmetry and, therefore, possesses a pair-breaking potential. 
b) This interaction induces pairs of bound states symmetric to $E_{\text{F}}$ within the gap of the quasiparticle excitation spectrum. 
For strong coupling ($k_\text{B} T_\text{K} \gtrsim  \Delta$) additionally a Kondo resonance occurs outside the excitation gap.}
\label{fig:sketch}
\end{figure}

\begin{figure}[t]
\includegraphics[width=1\columnwidth]{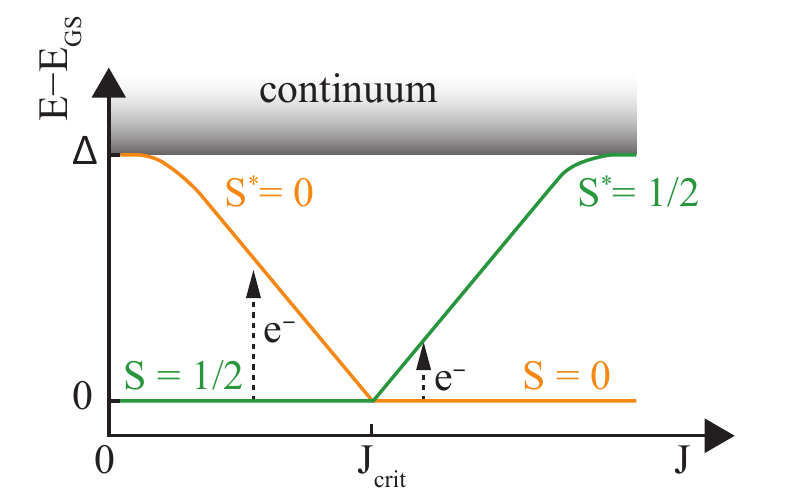}
\caption{Quasiparticle excitation diagram for a classical spin $1/2$ impurity on a superconductor. The exchange coupling $J$ with a local spin induces a low-lying state into the excitation gap of the superconductor. With an increasing exchange coupling $J$, a quantum phase transition changes
 the fermion parity of the ground state as the many-body spin of the total system changes from $S=1/2$ to $S=0$. 
}
\label{fig:diagram}
\end{figure}

\begin{figure}[t]
\includegraphics[width=1\columnwidth]{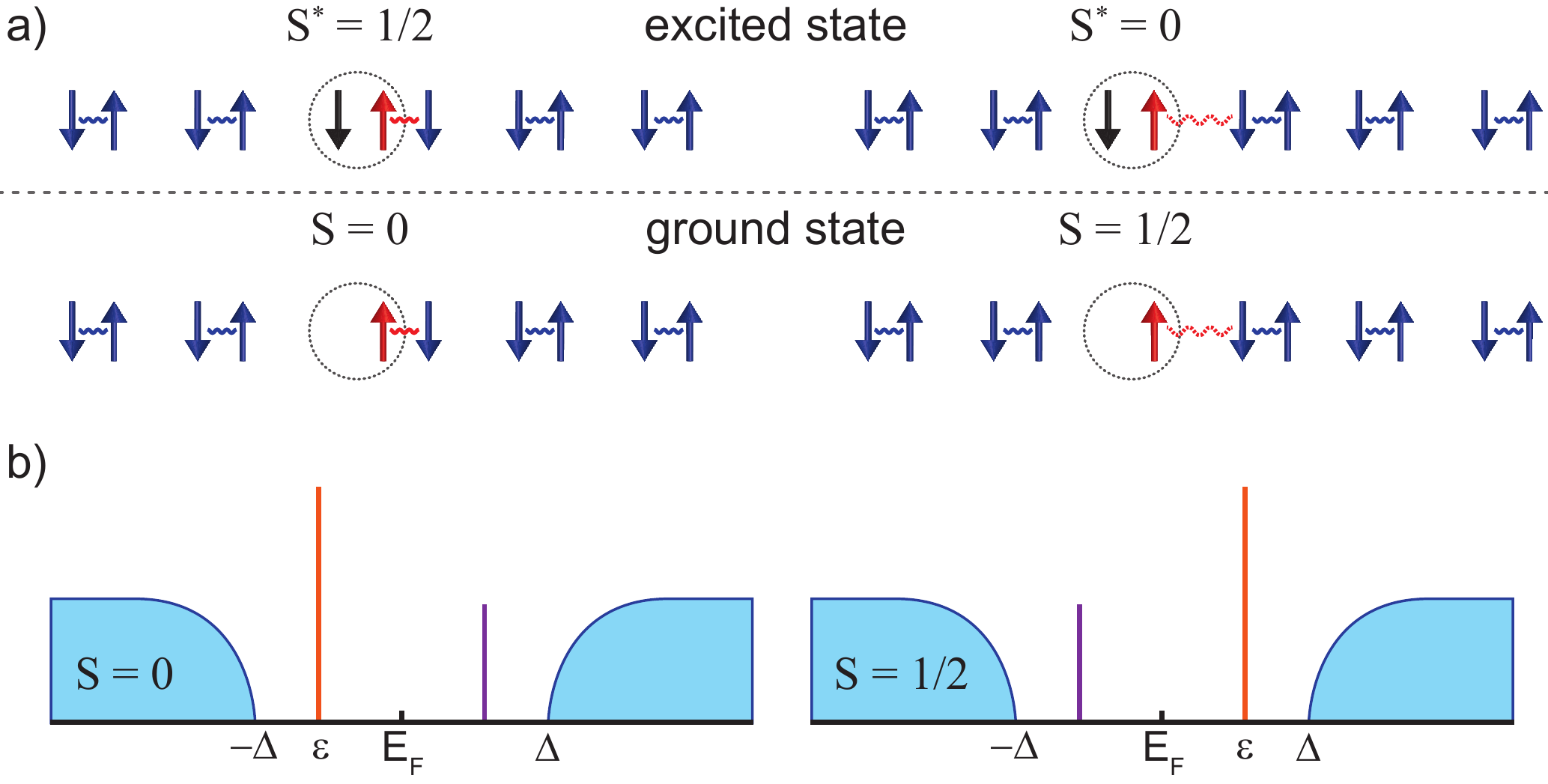}
\caption{a) Sketch of the two coupling regimes of a spin $1/2$ impurity on an $s$-wave superconductor: in the Kondo-screened case ($k_\text{B} T_\text{K} \gtrsim \Delta$), the coherent many-body ground state has zero net spin ($S=0$). The attachment of a tunneling electron then increases the spin to $1/2$ in the excited state ($S^*=1/2$). In the free-spin case ($k_\text{B} T_\text{K} \lesssim \Delta$), the screening of the local spin is incomplete and the many-body ground state spin is $S=1/2$. Here, the attachment of a tunneling electron induces the transition into an excited state with $S^*=0$.  b) The quantum phase transition from one to the other ground state by, {\it e.g.}, changing $J$, is accompanied by an inversion of the relative spectral weight of the electron and hole-like components of the bound state pair. This is caused by a crossing of $u$ and $v$, which changes their excitation character between particle- and hole-like.}
\label{fig:GS}
\end{figure}

Figure~\ref{fig:sketch}a presents a sketch of the important energy scales of the single-impurity problem neglecting the Coulomb scattering: the exchange coupling $J$ with the condensate competes with the superconducting pairing energy $\Delta$. The interaction of the local spin with the Cooper pairs then gives rise to a low-lying excited state within the gap of the quasiparticle excitation spectrum.

\subsubsection{Yu-Shiba-Rusinov states in the classical spin model}
Classical spin excitations were deduced independently by Luh Yu~\cite{Yu1965}, Hiroyuki Shiba~\cite{Shiba1968}, and A. I. Rusinov~\cite{Rusinov1968,Rusinov1969}.   The induced \YSR bound state
is a low-lying excitation of the many-body state within the excitation gap of the superconductor  (Fig.~\ref{fig:diagram}). 
In the simplest case of a spin $1/2$ impurity, a \YSR bound state gives rise to a pair of resonances symmetric in energy with respect to the Fermi level ($E_\mathrm{F}$) in the excitation spectra [Fig.~\ref{fig:sketch}\,(b)]~\cite{Zittartz70}. These resonances correspond to a quasiparticle excitation from the ground to the first excited state, which exists within the gap of the excitation spectrum.  
All other excitations of the system lie within the continuum of excitations outside the gap. In general, this first excitation changes the total spin by $\Delta S=\pm 1/2$. The excitation thus transfers either a singlet ground state, in which an electron from the bulk superconductor is bound to the impurity spin, to an excited doublet state, or vice versa  (Fig.~\ref{fig:GS}a). The nature of the ground state is determined by the coupling strength $J$. Note that the excitation of the superconducting ground state does not change the orbital occupation of the impurity. For an infinite system, it is irrelevant whether this excitation occurs via the particle- or hole-like component of the state, {\it i.e.}, via adding or extracting an electron from the system~\cite{Balatsky2006}.  Hence, bias-symmetric subgap resonances are a characteristic of   YSR excitations in tunneling spectra.

The energy of the bound state has been derived from a Bogoliubov transformation of the Hamiltonian of the combined impurity--BCS-superconductor system~\cite{Yu1965,Rusinov1968}, or via a Green's function
formulation, where the classical single-spin problem is solved via the T matrix~\cite{Shiba1968,Rusinov1969}. 
The bound state possesses an energy:
\begin{equation}
\varepsilon=\Delta \,\frac{1-a^2}{1+a^2},
\label{eq:YSRenergy}
\end{equation}

where $a=JS_{imp}\pi \rho_s$ (neglecting Coulomb scattering), with $\rho_s$ being the DoS at $E_\mathrm{F}$ in the normal state. 
Hence, the energy alignment of the \YSR resonance within the superconducting energy gap is determined to leading order by the exchange coupling strength $J$.

\subsubsection{Yu-Shiba-Rusinov states in the quantum spin model}
In the case of a quantum spin with antiferromagnetic exchange coupling to the quasi-particle reservoir, Kondo screening occurs on the superconductor as known for spins in normal metals~\cite{Kondo1964}. The formation of the Kondo singlet state then competes with the superconducting ground state~\cite{matsuura77}. For strong coupling, {\it i.e.}, $k_\mathrm{B} T_\mathrm{K} \gg \Delta$, a correlated Kondo state is formed, which screens the impurity spin and reduces the total spin to zero. Here,  $k_\mathrm{B}$ is the Boltzmann constant and  $T_\mathrm{K}$ the Kondo temperature.
In tunneling spectroscopy, a Kondo resonance is detected with a width of approximately $k_\mathrm{B} T_\mathrm{K}$ (Fig.~\ref{fig:sketch}c), while the bound states merge with the gap edge. 
In the case of very small coupling, {\it i.e.}, $k_\mathrm{B} T_\mathrm{K}\ll \Delta$, no screening occurs because the opening of the superconducting gap depletes the DoS on the Kondo energy scale $k_\mathrm{B} T_\mathrm{K}$ around $E_\mathrm{F}$. Again the bound states lie exponentially close to the gap edge.  In this case, the total spin of the system is $S>0$.\footnote{We restrict our discussion to the most interesting case, {\it i.e.}, to antiferromagnetic exchange coupling. For the ferromagnetic case, see, {\it e.g.}, Balatsky {\it et al.}~\cite{Balatsky2006}.
} 
The problem of a quantum spin was first solved by Matsuura~\cite{matsuura77}.  
The energy of the \YSR state depends on the  Kondo temperature $T_K$ and can be calculated, in the limit of $k_\mathrm{B} T_\mathrm{K}\gg \Delta$, using Eq.~\ref{eq:YSRenergy} 
and 
 \begin{equation}
a\approx \frac{\pi \Delta}{4\,k_\mathrm{B} T_\mathrm{K}}\ln{\frac{4\,k_\mathrm{B} T_\mathrm{K}}{\pi \Delta}e}.
\label{eq:alpha}
\end{equation}

An interesting situation arises, when both involved energy scales are similar: $k_\mathrm{B} T_\mathrm{K}\sim \Delta$. Then, the \YSR excitations lie well within the excitation gap of the superconductor (Fig.~\ref{fig:GS}b). Similar to the classical spin, the \YSR resonances correspond to a quasiparticle excitation from the ground to the first excited state. Again, the excitation changes the total spin by $\Delta S=\pm 1/2$. Here, it either excites the system from the free-spin ground state to the Kondo-screened excited state or from the Kondo-screened ground state to the free-spin excited state. The nature of the ground spin depends again on the coupling strength $J$. At $J_\mathrm{crit}$ both ground state levels cross. The level crossing signifies a quantum phase transition  between the free-spin state for weak coupling and the Kondo-screened state for strong coupling. It occurs at $k_\mathrm{B} T_\mathrm{K} \approx 0.3\,\Delta$ as calculated by numerical renormalization group theory~\cite{Satori1992,Sakai1993} (Fig.~\ref{fig:diagram}). At the point of the quantum phase transition  the fermion parity changes from even to odd. 
The spectral weight  of the particle- and hole-like excitation is influenced by the Coulomb potential~\cite{Balatsky2006}, which breaks particle-hole symmetry, and by asymmetries in the normal state conductance of the superconductor~\cite{Flatte1997PRL,Flatte97PRB, bauer2013}. The quantum phase transition has been treated in depth theoretically~\cite{Salkola1997,Bauer2007,Balatsky2006,matsuura77,Satori1992,Sakai1993,bauer2013,Sakurai70}. 

It is interesting to note that in the case of a quantum spin, only antiferromagnetic coupling to the substrate's reservoir yields the described quantum phase transition, while ferromagnetic coupling would yield weakly coupled bound states close to the gap edge~\cite{Satori1992}.

\subsubsection{Yu-Shiba-Rusinov wave function}

Now, we want to consider the wave functions of the \YSR state and their spatial extend. The \YSR wave functions can be determined after diagonalisation of the Hamiltonian of the combined BCS--impurity system by a Bogoliubov transformation. Solving the Bogoliubov-de Gennes equations yields the wave function $u$ and $v$, which are used to describe the superposition of electron- and hole-like excitations.
For the case of a point-like scatterer in a three-dimensional (3D) superconductor with an isotropic Fermi surface, Rusinov derived the wave functions as a function of distance $r$ as follows~\cite{Rusinov1968}: 
\begin{equation}
u(r), v(r) \propto \frac{\sin{\left({k_\mathrm{F} r + \delta^\pm}\right)}}{k_\mathrm{F} r}~\exp{\left[-\left|\sin{\left(\delta^+ - \delta^-\right)}\right| \frac{r}{\xi}\right]} ~.
\label{eq:ShibaWF}
\end{equation}
Both, $u$ and $v$ oscillate with $k_\mathrm{F} r$ (the Fermi wave vector times the distance from the impurity), but show a different scattering phase shift $\delta^\pm$. Interestingly, the energy of the bound state is related to the difference in the scattering phase shift for $u$ and $v$:
\begin{equation}
\epsilon=\Delta \cos(\delta^+ - \delta^-).
\label{eq:shift}
\end{equation}
 The wave functions' fall-off away from the impurity is dominated by the factor $1/({k_\mathrm{F} r})$ for short distances. But for large distances, the exponential decay dominates, which depends on the  superconducting coherence length  $\xi$. $\left|u(r)\right|^2$ and $\left|v(r)\right|^2$ can be probed, \eg, at positive and negative sample bias in a scanning tunneling microscope. However, we note that the character of $u$ and $v$, \ie, whether it is particle-like or hole-like, changes when undergoing the quantum phase transition. We will discuss this further in Section~\ref{Sec:QP}.

\begin{figure*}[t]
\includegraphics[width=1.5\columnwidth]{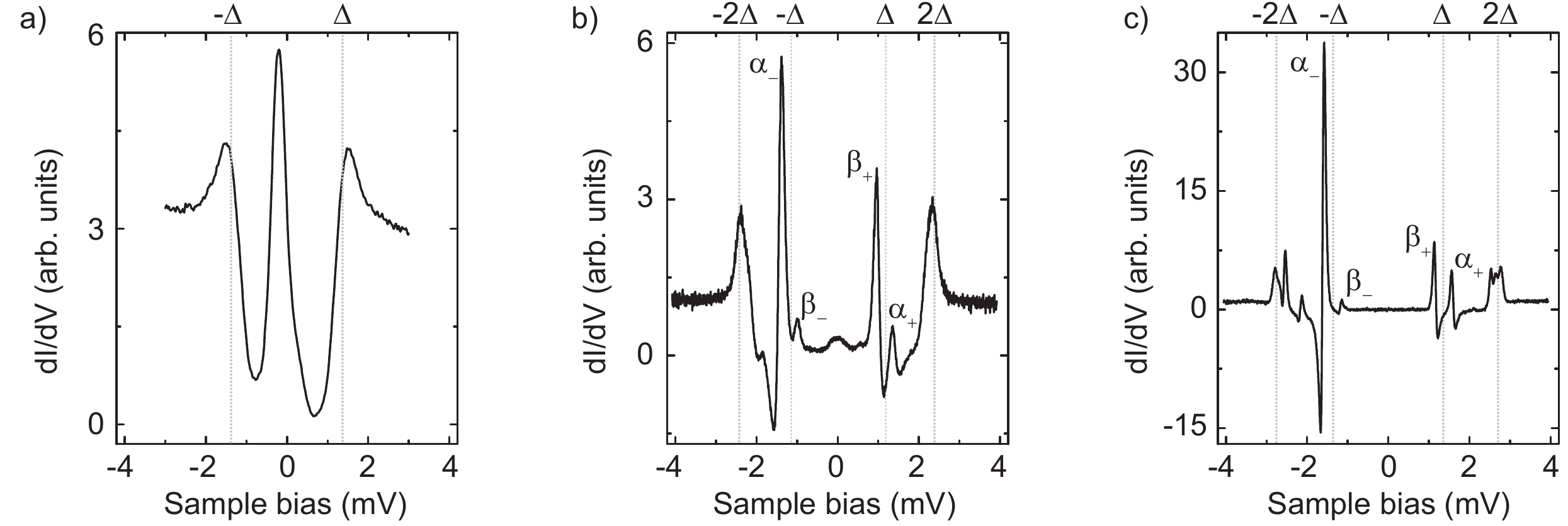}
\caption{Tunneling spectra of Mn adatoms on Pb(111) acquired with a metal tip at 1.2\,K (a), or with a superconducting Pb-covered tip at 4.8\,K (b), and 1.2\,K (c), respectively. Different numbers of \YSR resonances are resolved depending on the tip material and the temperature. 
Figure adapted from Ref.~\cite{ruby2015}.}
\label{fig:tip}
\end{figure*}

\section{Transport through Yu-Shiba-Rusinov states}
\label{Detection}
As described above, \YSR states occur as resonance in an excitation spectrum. Such a spectrum can be recorded in different geometries and with different tunnel coupling strengths. The best known examples are tunneling spectroscopy and transport through quantum dots.

\subsection{Resolving Yu-Shiba-Rusinov states in tunneling spectroscopy}
\label{STS}

In a tunnel junction, the above mentioned excitation is induced by particles (holes) tunneling between the two electrodes via the bound state. Hence, $u$ and $v$ are probed in spectra of the differential conductance $dI/dV$ as resonances at opposite bias with respect to $E_\mathrm{F}$ within the superconducting gap. As a typical example, \Fig{tip} shows  $dI/dV$ spectra of a Mn adatom on Pb(111).
 
\begin{figure*}[t]
\includegraphics[width=1.5\columnwidth]{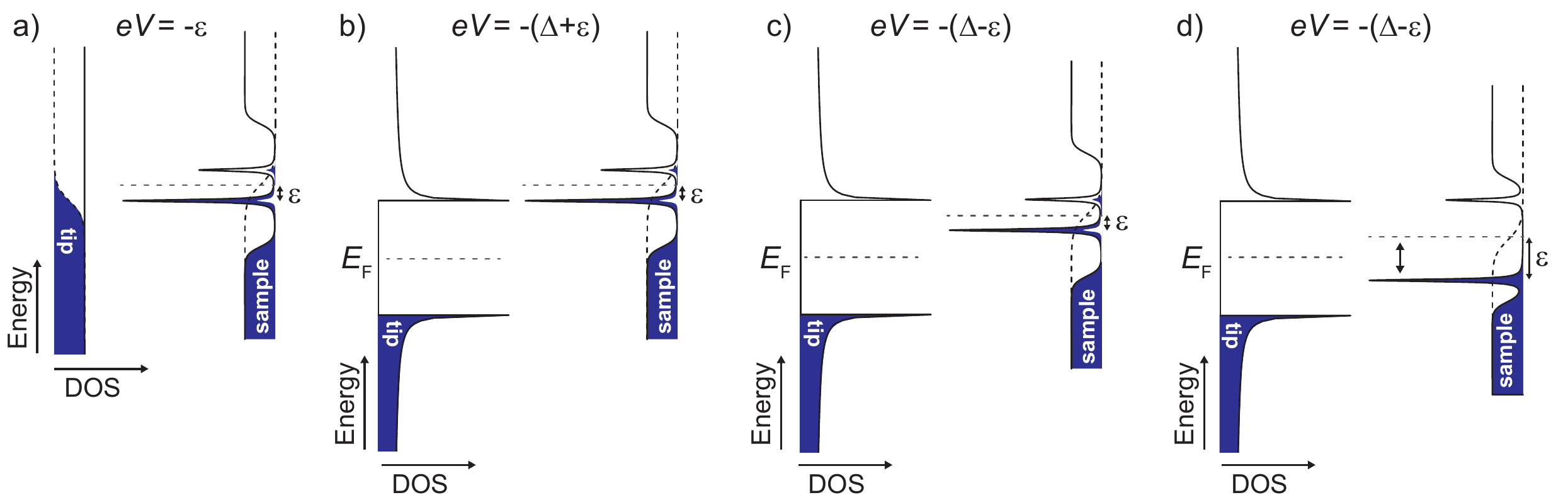}
\caption{
Scheme of the tunneling between a \YSR resonance in the sample and the tip DoS with a metallic (a) and a superconducting tip (b -- d). 
a) Tunneling between the hole-like component of the \YSR state and the tip sets in at a sample bias of $eV=-\epsilon$. The energy resolution is limited by the thermal broadening of the tip's Fermi edge. 
 b) At  $eV=-(\DeltaT+\epsilon)$, tunneling occurs between the hole-like  component of the \YSR state and the particle-like BCS quasiparticle resonance of the tip.
c) At finite temperature, thermally activated tunneling can also occur at $eV=-(\DeltaT-\epsilon)$ if the bound state energy $\epsilon \lesssim k_\mathrm{B} T$. 
d) If $\epsilon \gtrsim k_\mathrm{B} T$, no thermal tunneling is observed.   Figure adapted from Ref.~\cite{hatter2015}.}
\label{fig:STS_SC}
\end{figure*}

If a metallic tip is used as second electrode (spectrum in \Fig{tip}a), tunneling into the electron and hole component of the bound state sets in at a sample bias of $eV=\pm\epsilon$ as sketched in \Fig{STS_SC}a. Tunneling into the excitation continuum occurs only for $|eV|\geq\DeltaS$. The energy resolution (usually) is limited by the temperature-dependent Fermi-Dirac broadening of the Fermi edge of the tip ($\approx 300\,\mu V$ at 1.2\,K), which yields sizeable smearing of the \YSR resonances. To illustrate this broadening, we show in \Fig{tip}a a \didv spectrum recorded on a Mn atom on Pb(111). We resolve only a single \YSR resonance at negative bias. At positive bias, a shoulder appears, which is linked to the particle component $u$ of the \YSR state, but cannot be resolved unambiguously.  

In order to circumvent the thermal limit of the energy resolution, superconducting tips were used by several groups~\cite{Ji2008,Ji2010a,Ternes2006,Franke2011,heinrich2013,ruby2015a}. These tips probe the sample DoS with the sharp features of the BCS-like tip DoS instead of a temperature-broadened Fermi edge of a metal tip (see \Fig{STS_SC}b). This yields  a considerable gain in energy resolution already at 4.8\,K (\Fig{tip}b).
Yet, the spectrum is a convolution of the sharp features in the tip and sample DoS: the spectral function of the sample is probed by the electron and hole quasiparticle resonances at $+\DeltaT$ and $-\DeltaT$, respectively ($\DeltaT$ is the energy of the gap parameter of the tip). Hence, the spectral function of the sample appears shifted by $\pm\DeltaT$. A subgap resonance with energy $\epsilon$ ({\it e.g.}, $\alpha_\pm$ in \Fig{tip}b) appears at  $eV=\pm(\DeltaT+\epsilon)$ , {\it i.e.}, when the particle-like singularity  in the tip's spectral function is aligned with the hole component of the \YSR excitation and vice versa (see \Fig{STS_SC}b for a sketch of the tunneling configuration).

Additionally, thermally activated tunneling~\cite{Franke2011,ruby2015,hatter2015,ruby2016} 
occurs at finite temperature. The thermal energy can induce an excitation of the \YSR state. Hence, there is a finite probability, which scales with $\varepsilon$ and $T$, to find the system in the excited state. This "thermal" population is probed at $eV=\pm(\DeltaT-\epsilon)$, {\it i.e.}, when the hole(electron)-like component of the \YSR excitation is aligned with the hole(electron)-like singularity in the tip spectral function (\Fig{STS_SC}c). 
This gives rise to the {\it thermal} resonances $\beta_\pm$ in \Fig{tip}b, which are (thermal) {\it replicas} of resonances $\alpha_\mp$.

In order to extract the actual spectral function of the sample, a deconvolution of the $dI/dV$ spectra is possible. However, this requires knowledge of the tip spectral function, which can usually be inferred from spectra of the pristine superconductor. Then, either a direct deconvolution \cite{Choi16}, or a fit of convoluted spectral functions to the $dI/dV$ \cite{hatter2015} can be performed.

 A further increase in energy resolution is achieved  by lowering the experimental temperature. The $dI/dV$ spectrum in \Figure{tip}c was acquired with a superconducting tip at $1.2\,$K. The spectrum resolves particle and hole excitations of three \YSR states. The line width is reduced compared to the spectrum acquired at $4.8\,$K. The energy resolution can be as good as $\delta E\approx 50\umueV$ at $1.1\,$K~\cite{ruby2015a}. 
Interestingly, thermal resonances are only observed for the \YSR excitation closest to $E_\mathrm{F}$ ($\beta_\pm$). 
The probability of a thermal excitation of a \YSR state (and, hence, the intensity of the thermal resonance in $dI/dV$) depends on the ratio of the excitation energy $\epsilon$ and the thermal energy $k_\mathrm{B} T$.  
The larger $\epsilon$, {\it i.e.}, the further the \YSR state is away from $E_\mathrm{F}$, the smaller is its thermal population (comp. \Fig{STS_SC}c and \ref{fig:STS_SC}d). While at 4.8\,K all \YSR resonances posses thermal counterparts, at 1.2\,K, this is limited to excitations close to $E_\mathrm{F}$. It is noteworthy that at 4.8\,K, there is also a sizeable amount of    thermally excited quasiparticles in the pristine superconductor (sample and tip).

\begin{figure}[b]
\includegraphics[width=1\columnwidth]{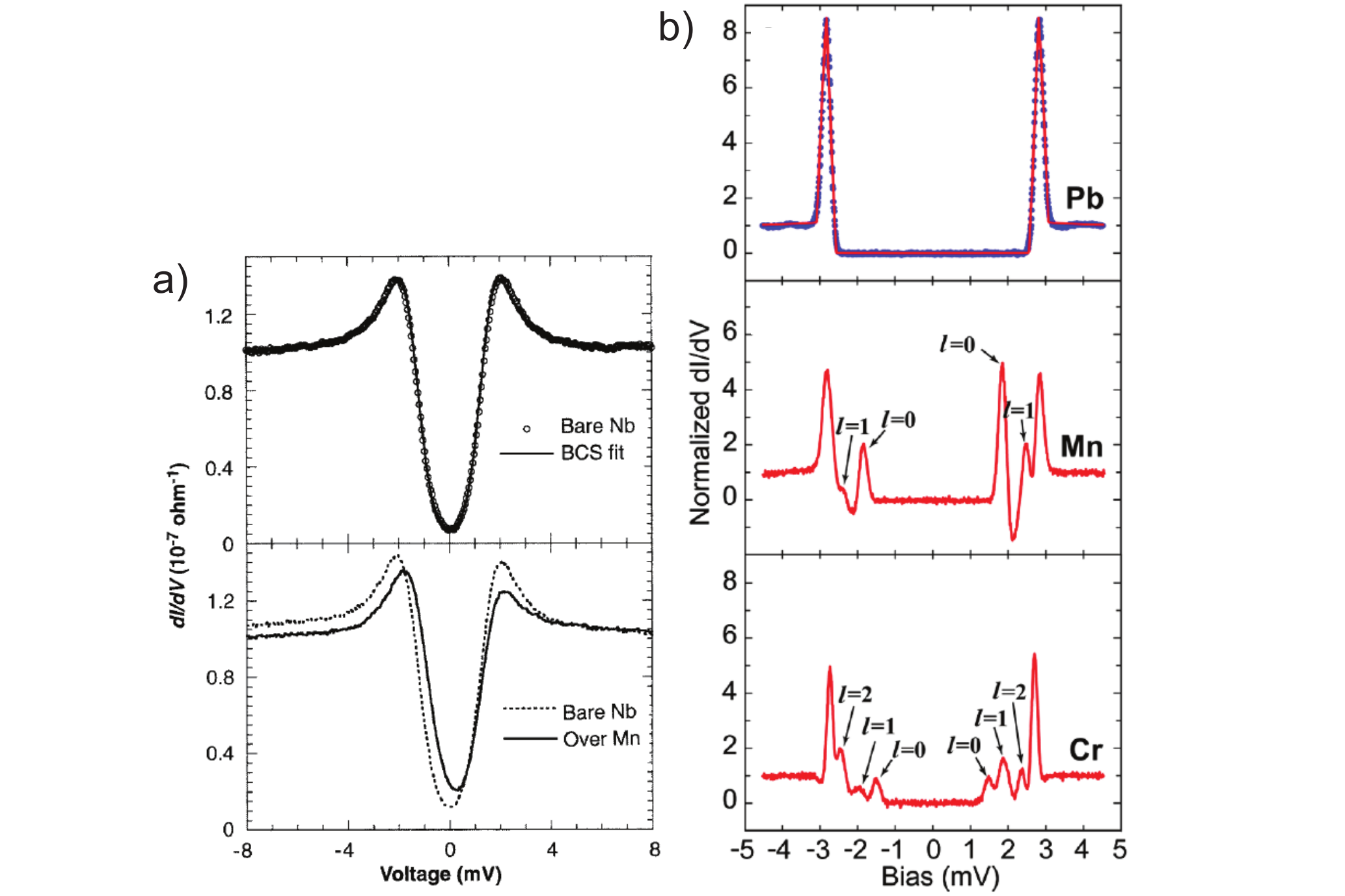}
\caption{a) $dI/dV$ spectra over superconducting Nb(110) and over a Mn adatom on Nb measured at $3.85$\,K with a metallic tip. From Ref.~\cite{Yazdani1997}. Reprint with permission from AAAS.  
b) Mn and Cr adatoms on Pb films present multiple Shiba resonances in the gap. $dI/dV$ measured with a superconducting Nb tip at $0.4$\,K. Reprinted figure with permission from Ref.~\cite{Ji2008}. Copyright 2008 by the American Physical Society.
}
\label{fig:lit}
\end{figure}

Historically, the investigation of single impurities became possible when ultra-high vacuum (UHV) preparation techniques were combined with low temperatures. In 1997, A. Yazdani and co-workers could, for the first time, access \YSR states induced by single paramagnetic impurities with an UHV STM operating at $3.8$\,K~\cite{Yazdani1997} (Figure~\ref{fig:lit}a). Furthermore, the lateral resolution of the microscope enabled the detection of the wave function fall-off of the \YSR state, which happens on atomic distances on 3D superconductors. These experiments provided the first magnetic fingerprint on a single atom resolved by low temperature STM, just before the first evidence of Kondo resonances in 1998~\cite{Madhavan98,Li98}, and prior to the detection of inelastic excitations of spin eigenstates in 2004~\cite{Heinrich04}, or of a spin-polarized conduction signal in 2007~\cite{Yayon2007}. 

Ten years later and with improved energy resolution, multiple Shiba resonances of single Mn and Chromium (Cr) adatoms on Pb films could be resolved using a superconducting Nb tip and lowering the temperature to $0.4$\,K by Ji {\it et al.}~\cite{Ji2008} (Figure~\ref{fig:lit}b). 
Since then, the field has progressed rapidly. Several groups have focused on resolving the quantum many-body ground state \cite{Franke2011,hatter2015,Hatter2017}, peculiarities in the spatial decay of the YSR wave functions \cite{ruby2016, Choi16,Menard2015}, the interaction of impurities \cite{Kezilebieke2017,ChoiArxiv2017}, or different transport processes through the \YSR states \cite{ruby2015, Randeria2016}. In all these investigations, the substrate are $s$-wave superconductors. We will concentrate on these in this review.

However, we need to mention that even prior to these works, several important results were obtained on subgap resonances in Bi$_2$Sr$_2$CaCu$_2$O$_{8+\delta}$, a high-T$_c$ cuprate superconductors~\cite{Hudson1999,Yazdani1999a,Pan2000,Hudson2001}. Yazdani and coworkers~\cite{Yazdani1999a} showed that intrinsic defects as well as non-magnetic surface adsorbats induce subgap bound states in the excitation gap of this $d$-wave superconductor. Hudson {\it et al.} finally compared the effect of nominally non-magnetic Zn dopands with paramagnetic Ni impurities~\cite{Hudson2001}.  In contrast to $s$-wave superconductors, in a $d$-wave substrate, also these nonmagnetic impurities posses a pair-breaking potential. A more detailed summary of these results can be found in the review of Balatsky, Vekhter, and Zhu~\cite{Balatsky2006} and is beyond the scope of this review.

\subsection{Resonantly enhanced Andreev reflections through Yu-Shiba-Rusinov states }

As described in the previous section, excitations of $u$ and $v$ appear as resonances symmetric to $E_\mathrm{F}$ in tunneling spectra of the differential conductance $dI/dV$. Single-particle tunneling into/ out off  these subgap states excites the impurity--superconductor system from its many-body ground state to the first excited state, which is of a different fermion parity. It is important to note that a continuous single particle current requires an efficient relaxation mechanism of the excitation (\eg, $\Gamma_1$ in Fig.~\ref{fig:transport}c), with a rate faster than the tunneling rate. 
An intriguing transport mechanism that comes into play when using at least one superconducting electrode is the tunneling of an electron through the barrier and its concomitant retroreflection as a hole. This so-called Andreev reflection transfers a Cooper pair, {\it i.e.}, two particles, from one electrode to the other. Usually, Andreev processes between the quasiparticle continua of tip and sample are only important at strong tunneling coupling~\cite{Ternes2006}. This is easily understood considering that the tunneling barrier has to be overcome twice, by an electron and by its reflected hole. However, \YSR  states resonantly enhance the Andreev transport~\cite{Martin2014,Andersen2011,Ioselevich2013}. Then, a Cooper pair is transferred into/ out off the condensate via both $u$ and $v$ and the state occupation is unchanged (Fig.~\ref{fig:transport}c, right panel). Hence, the many-body ground state is conserved and no relaxation is required for a continuous current.  Both transport mechanisms occur with the same bias threshold  and it is a priori ambiguous, which of the two is dominating.  It is noteworthy that also multiple Andreev reflections via \YSR states have been observed in STS experiment~\cite{Randeria2016}. These involve the transfer of more than two particles (holes), but occur with different bias thresholds.

Beside single atomic impurities, also quantum dots with odd filling can induce \YSR states when they are coupled to at least one superconducting lead. 
These states can be accessed by transport experiments~\cite{Deacon2010b,DeFranceschi2010,Pillet2010,Dirks2010,Chang2013,Lee2014,Zanten2015,Jellinggaard2016,Lee2017}. By means of gate potentials, one can tune the energy of the bound state and drive the singlet-doublet phase transition of the quantum ground state~\cite{Deacon2010b,Lee2017}. 

These transport measurements are typically interpreted in terms of Andreev processes due to the strong tunneling coupling strength~\cite{Deacon2010b,Pillet2010,Dirks2010,Lee2014}. They are traditionally referred to as Andreev bound states -- as opposed to \YSR (or Shiba) states, the denominations commonly used in the STM community. On the contrary, $dI/dV$ spectra in scanning tunneling experiments are usually directly linked to the weight of $u$ and $v$ of the \YSR wave function~\cite{Yazdani1997,Franke2011}. However, this assumption holds only for single-particle tunneling. 
\vspace{1.0 cm}

\subsection{From single-particle tunneling to Andreev transport}
\label{Sec:transport}

\begin{figure}[t]
\includegraphics[width=1\columnwidth]{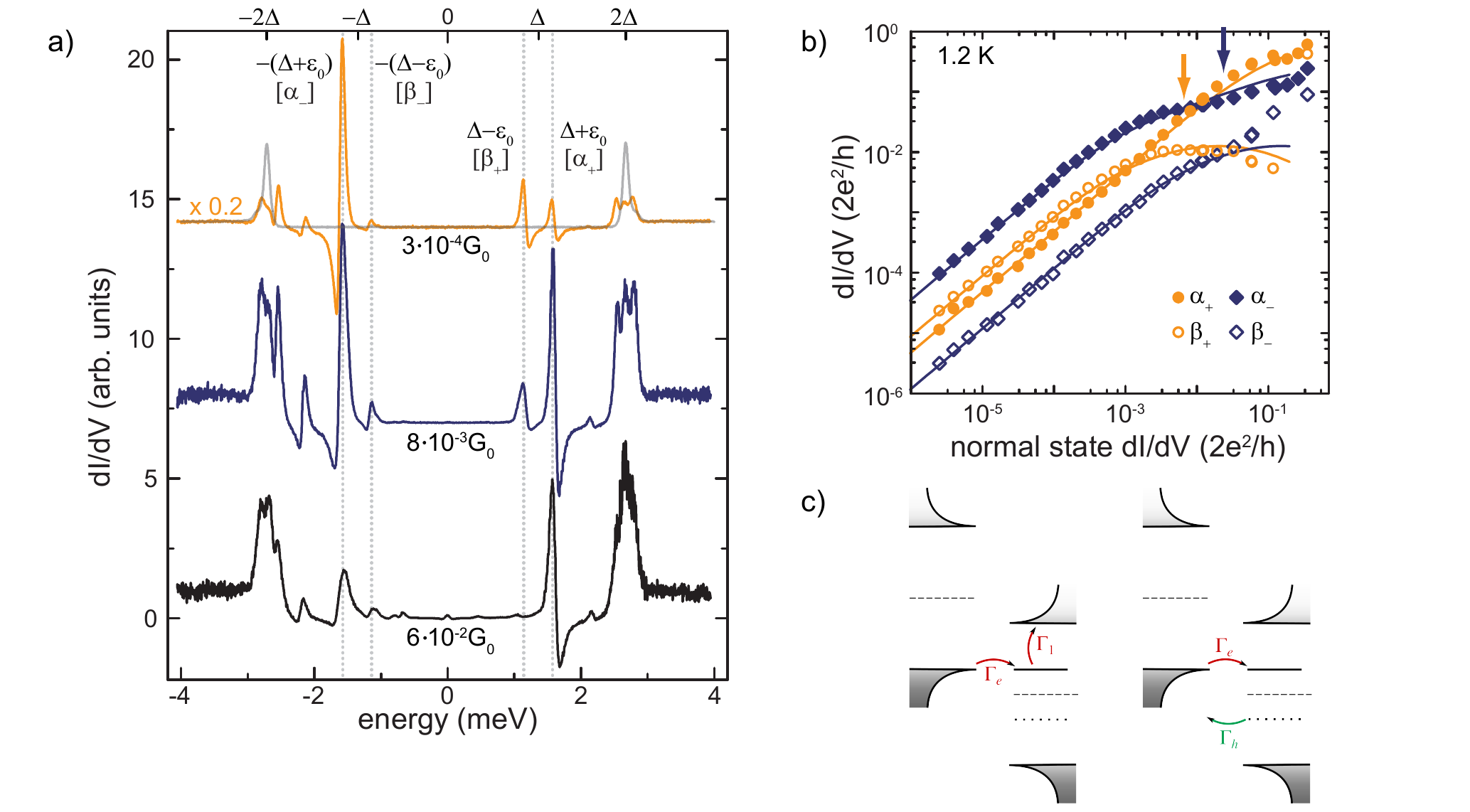}
\caption{Mn adatom on Pb(111). a) $dI/dV$ spectra over the adatom at different tip-sample distances acquired at $1.2\uK$. The relative weight of  $+\alpha$ and $-\alpha$ in the $dI/dV$ spectra above the adatom change with the junction resistance.
b) Intensity of $\pm\alpha$ and $\pm\beta$ as a function of junction resistance. The arrows indicate the turnover from single-particle dominated transport to Andreev dominated transport. 
c) Sketches of the  transport processes through an isolated level in the excitation gap. Single particle transport (left) requires a relaxation to the continuum, while an Andreev reflection transfers a Cooper pair but does not change the occupation of the level.  
Figure adapted from Ref.~\cite{ruby2015}.}
\label{fig:transport}
\end{figure}

In order to shed light on the different  contributions to tunneling, in Ref.~\cite{ruby2015}, the excitation spectrum on Mn adatoms on Pb(111) was probed as a function of junction resistance. In Figure~\ref{fig:transport}a a change in the relative weight of resonances $+\alpha$ and $-\alpha$ is observed when measuring the $dI/dV$ with a superconducting Pb tip at different tip--sample distances, \ie, with decreasing junction resistance from top to bottom. \Figure{transport}b presents a quantitative analysis over five orders of magnitude of normal state junction resistance of this \YSR excitation and its thermal counterpart ($\pm\beta$). At large tip--sample distances (low conductance), a linear increase of all intensities is observed over several orders of magnitude of normal state conductance.
At higher conductance, a sublinear behavior is detected and a crossover of the intensities of the electron- and hole-like part is observed for both, the \YSR resonances ($\pm\alpha$) and their thermal counterparts ($\pm\beta$).

Qualitatively, this  is understood considering the transport as sketched in \Fig{transport}c. A single-particle current at positive bias is proportional to the product of the relaxation rate $\Gamma_1$ and the tunneling rate $\Gamma_e$, which in turn scales with the weight of the particle-like component~\cite{ruby2015}.  It is independent of the hole-like component. Consequently, the intensities in $dI/dV$ spectra at positive and negative bias voltage [$eV=\pm(\DeltaT+\epsilon)$] reflect the relative weight of $u$ and $v$. As long as $\Gamma_e$ is small compared to $\Gamma_1$, the single particle current increases linearly with the normal state conductance. When the tunneling rate $\Gamma_e$ becomes similar or larger than the relaxation rate $\Gamma_1$, then the relaxation is limiting the single particle current and yields a sublinear increase compared to the normal state conductance. 

Andreev tunneling, on the contrary, proceeds via both, $u$ and $v$. The tunneling rate is then the convolution of $\Gamma_e$ and $\Gamma_h$. It is independent of the relaxation rate $\Gamma_1$. Therefore, this process becomes important for strong tunneling couplings, \ie, for large currents. Measured with a superconducting tip, the relative intensity of the hole- and electron-like resonances invert in the spectra at large tunnel coupling. It is noteworthy that, with a metallic tip, Andreev transport yields the same intensities for both resonances, \ie, at positive and negative bias~\cite{Martin2014}, while single-particle tunneling again reflects the asymmetry of the $u$ and $v$ component in the wave function~\cite{ruby2015}. 

A standard Keldysh calculation of the tunneling current, which used all tunneling rates as well as phenomenological relaxation rates of the excited state ($\Gamma_1$ and $\Gamma_2$),  yielded the fits to the conductance evolution shown in \Fig{transport}b. The arrows indicate the crossover between single-particle and Andreev-dominated transport. From the fits, the inverse relaxation rates were determined to be in the order of hundreds of picoseconds at $1.2\uK$. However, it was found that they decreased to $\approx6\,$ps at a temperature to $4.8\uK$, which indicates a temperature-driven relaxation process. The analysis, which was based on a single step relaxation as sketched in \Fig{transport}c, overestimates the temperature-dependence by two orders of magnitude. This discrepancy with the experimental observation was understood by taking into account the presence of additional \YSR excitations in the spectrum, which provide additional, more efficient relaxation channels  involving more than one \YSR resonance and the formation of Cooper pairs. 

The measurements show that tunneling experiments can be used to extract quantitative information on the electron and hole component of the \YSR states, as well as on the relaxation rates.

\section{Orbital character of Yu-Shiba-Rusinov states}
\label{Sec:orbital}

In Section~\ref{Sec:transport}, we discussed the transport mechanism through \YSR resonances within the superconducting gap disregarding the number of \YSR resonances. In literature, several experiments were described, in which more than one pair of subgap resonances had been observed~\cite{Ji2008,Ji2010a,ruby2015,hatter2015,Randeria2016}. It was suggested that the coupling to other degrees of freedom, like magnetocrystalline anisotropy~\cite{Zitko2011}, or vibrations~\cite{Golez2012} can explain the experimental observation of several \YSR states~\cite{hatter2015}.
Before including possible additional excitations in the discussion (see Section~\ref{SubSec:anisotropy} below), we point out that a single impurity can bear multiple \YSR states due to the presence of several orbital scattering channels.

Early experiments~\cite{Ji2008,Ji2010a} linked multiple \YSR states to scattering channels with different angular momenta ($l=0,1,2, ...$)~\cite{Flatte1997PRL,Flatte97PRB}. Only in 2008, Moca and coworkers showed theoretically that, beside several scattering channels, the internal, \ie, orbital structure of the impurity plays a decisive role in the number and nature of \YSR states in the excitation spectrum~\cite{Moca2008}. Transition metal atoms carry their spin in form of unpaired electrons in the $d$ shell. The local environment, \ie, the crystal field experienced by the impurity (partially) lifts the degeneracy of the $d$ levels according to their symmetry with respect to the crystal field. This in turn yields an orbital-dependent potential and exchange coupling with the substrate, which is imprinted in the \YSR states.   

For example, Mn  adatoms on Pb are expected to  be in a $d^5$ configuration and host an unpaired electron in each of the five $d$ levels. Because of the $^{6}S_{5/2}$ nature of the ion, only conduction electrons with $l=2$ are scattered~\cite{Schrieffer1967}. This results in five \YSR states, some of which can still be degenerate because of the actual symmetry of the environment. 

\begin{figure}[t]
\includegraphics[width=1\columnwidth]{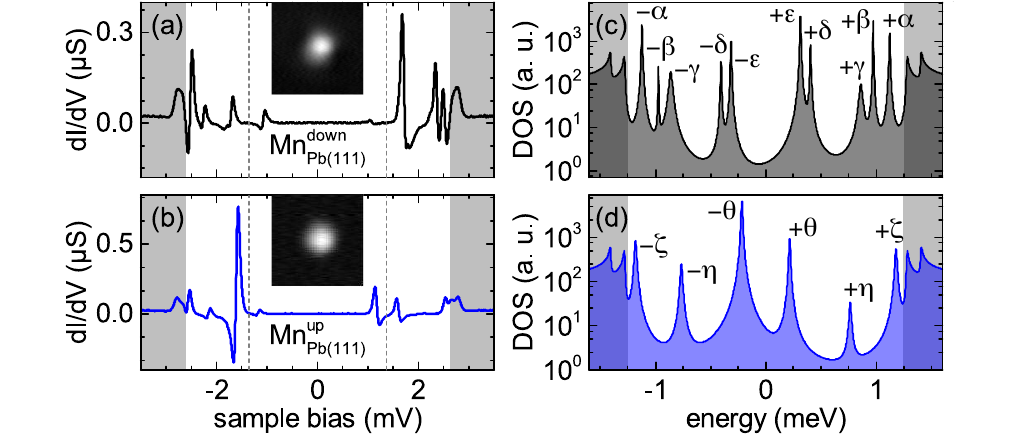}
\caption{Mn atoms in two different stable adsorption sites on Pb(111). (a,b) \didv spectra of the different Mn adatoms, labeled Mn$^\mathrm{up}$ and Mn$^\mathrm{down}$ according to their topographic appearance.
The shaded areas indicate the two BCS coherence peaks and excitation continuum, the dashed lines the tip gap ($\pm 1.38$\,mV). The insets show topographies of the adatoms. (c) Deconvolved sample DoS the spectrum in (a), which exhibits 
five \YSR resonances. (d) Deconvolved DoS of the spectrum in (b). 
There are three YSR resonances. Figure adapted from Ref.~\cite{ruby2016}.}
\label{fig:MnPb111}
\end{figure}

\begin{figure*}[t]
\includegraphics[width=1.5\columnwidth]{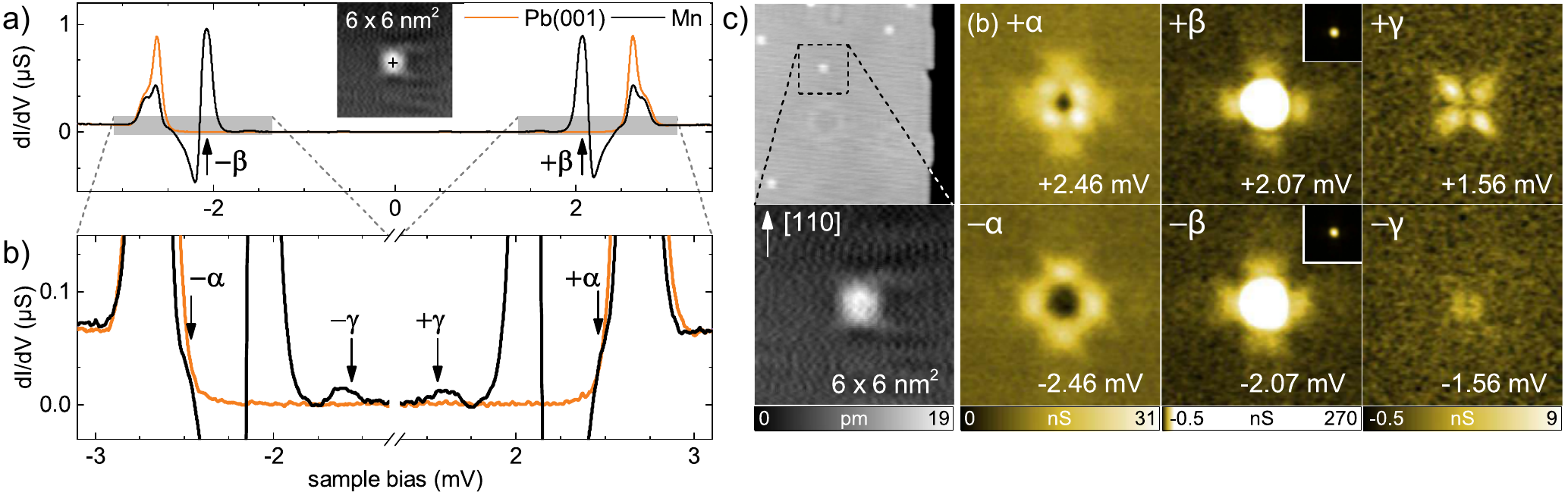}
\caption{Mn adatoms on Pb(001). a) and b) The $dI/dV$ spectrum above the adatom unveils three pairs of YSR resonances. c) Topography (left) and $dI/dV$ maps at the energies of the YSR resonances $\pm\alpha$, $\pm\beta$, and $\pm\gamma$. The maps reveal the fourfold symmetry of the $d$ levels and the surface. Figure adapted from Ref.~\cite{ruby2016}.}
\label{fig:orbital}
\end{figure*}
Mn adatoms on Pb(111) acquire two stable adsorption sites which differ in apparent height by $\approx 0.6$\,\AA~(\Fig{MnPb111})~\cite{ruby2016}. For the higher species with a symmetric appearance three pairs of \YSR resonances are observed (\Fig{MnPb111}b,d). This is in line with a threefold symmetric hollow adsorption site on the (111) lattice, which splits the $d$ levels into three subsets, the two-fold degenerate $d_{xz,yz}$ and $d_{xy,x^2-y^2}$, and the non-degenerate $d_{z^2}$. Each of these subsets is at the origin of one of the pairs of \YSR states. 
Yet, for the second stable adsorption site for adatoms on the (111) surface with an asymmetric topographic appearance and a lower apparent height (\Fig{MnPb111}a,c), \didv spectra exhibit five pairs of \YSR resonances, which implies a complete lifting of all $d$ level degeneracies caused by the absence of higher spatial symmetry~\cite{ruby2016} .  
Similarly, also for Cr adatoms on Pb(111), five \YSR states are observed (see below)~\cite{Choi16}.

A direct link between a single pair of  \YSR states and a certain $d$ level can be most easily derived if the symmetry of the surface matches the symmetry of the real space representation of the $d$ level. This is the case for a hollow-site adsorption on a fourfold-symmetric Pb(001) surface, which yields a square pyramidal coordination. This geometry lifts all degeneracies except the degeneracy of the $d_{xz}$ and $d_{yz}$. Yet, the $d_{xy}$ orbital is close in energy to (or even degenerate with) these  $d_{\pi}$ orbitals. The energy difference depends on the in-plane and out-of-plane nearest-neighbor distance. 

Mn adatoms on Pb(001) (\Fig{orbital}a,b) induce a pair of dominating resonances $\pm\beta$ and  two faint pairs of resonances ($\pm\alpha$ and $\pm\gamma$)~\cite{ruby2016}. This is in line with the above described coordination geometry assuming (almost) degenerate $d_{xy}$ and $d_{xz,yz}$. 
\Figure{orbital}c presents \didv maps at the energy of the three intragap resonances, which reveal that each YSR excitation presents a characteristic spatial pattern. $\pm\beta$ are most intense and mainly circular symmetric (note the stretched color code of the \didv maps). These \YSR resonances originate from the  $d_{z^2}$ orbital because this orbital has a C$_1$ symmetry and the strongest wave function overlap with the tip. The maps of $\pm\alpha$ and $\pm\gamma$ all exhibit fourfold symmetric patterns, which match the symmetries of the remaining $d$ levels. While $\pm\gamma$ is linked to the degenerate $d_{xz,yz}$ and $d_{xy}$ orbitals as can be inferred from a splitting of these resonances in the presence of variations in the local environment, the $d_{x^2-y^2}$ orbital is at the origin of resonances $\pm\alpha$~\cite{ruby2016}.

\begin{figure*}[t]
	\includegraphics[width=1.5\columnwidth]{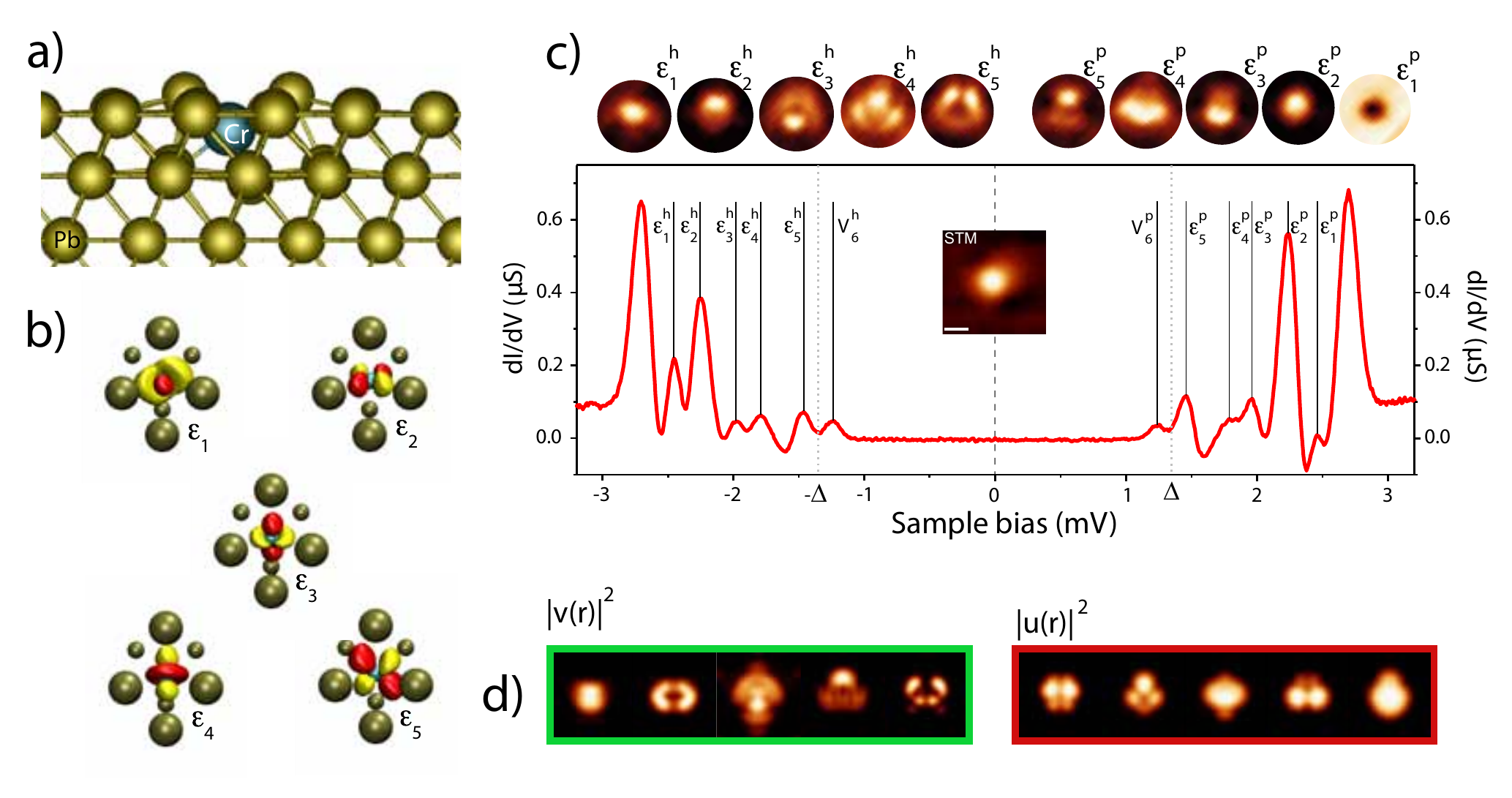}
	\caption{Cr atoms on Pb(111): (a) Results of a full DFT relaxation showing the most stable position of a Cr atom (blue) embedded under the surface of Pb(111). (b) Top-view of wave function  amplitude isosurfaces for the five spin-polarized orbitals of embedded Cr, including in yellow, the seven closest Pb atoms. These five states are the only ones originating from the original $d$ manifold of the Cr atom with relevant weight.  (c) \didv spectrum on a Cr atom. The five peaks labeled $\epsilon_\mathrm{n}^{(h,p)}$ represent the particle (p) and hole (h) excitations of the five YSR states. The peaks labeled $\nu_5^{(h,p)}$ are the thermal replica of $\epsilon_5^{(h,p)}$. The maps on top of the graph depict the amplitude of the peaks in a region of about 2\,nm diameter around the impurity. (d) Simulation of $|u(r)|^2$ and $|v(r)|^2$ for each scattering channel shown in (b). Figure adapted from Ref.~\cite{Choi16}.
		}
	\label{fig:Cr}
\end{figure*}

The studies performed on  Mn/Pb(111)  revealed the existence of two  atomic configurations, each with a different YSR fingerprint. 
A similar experimental study  on the  Cr/Pb(111) system was supported by Density Functional THeory (DFT) simulations~\cite{Choi16} which unveiled that a subsurface configuration with the impurity atom embedded underneath the top-most Pb layer was energetically preferred (Fig.~\ref{fig:Cr}a). 
DFT results determined that the barrier for Cr atoms to reach the subsurface site through hollow sites of the (111) surface is only 21\,meV,  which is smaller than the lateral diffusion barriers and can easily be overcome by "hot" adatoms before thermalisation. 
Subsurface Cr impurities show  spectra full of intra-gap features attributed to five \YSR excitation pairs (Fig. \ref{fig:Cr}c). With support of DFT calculations, these were interpreted as originating from the five half-filled $d$ orbitals of Cr, which act as spin-polarized scattering channels. Similar to the case of Mn atoms of lower apparent height (Mn$^\mathrm{down}$), the local symmetry around the embedded atom is reduced and, consequently, the degeneracy of the $d$ subshell  broken into a multiplet of five spin-polarized $d$ resonances.     

Although the DFT simulations reveal that the Cr  states are  mixed with Pb bands, the five spin-polarized channels maintain  some degree of  the atomic $d$ character (Fig. \ref{fig:Cr}b). Hence, in a first approximation, they can be treated as five independent channels with a different exchange interaction $J_\mathrm{n}$ [$\mathrm{n}=(1,...,5)$], producing  five YSR bound states, each with a different excitation energy $\epsilon_n$.  We can use the computed single-particle wave function of the five scattering channels for a calculation of the Bogoliubov quasiparticle coefficients  $|u_{\mathrm{n}} (\vect{r})|^2$ and $|v_{\mathrm{n}} (\vect{r})|^2$. These represent the local DoS of the particle and hole components. The spatial maps of these amplitudes (Fig. \ref{fig:Cr}d) show that, while the  $|u_{\mathrm{n}} (\vect{r})|^2$ component resemble closely the shape of the scattering orbital, the corresponding $|v_{\mathrm{n}} (\vect{r})|^2$ component  differs strongly. Such asymmetry in the shape of   Shiba components was also observed in the experimental maps (Fig. \ref{fig:Cr}c) of the particle- and hole-like \YSR excitations. As expected the  simulated orbital shapes can be recognized in some of the particle YSR maps (sample bias $>0$), while the corresponding hole maps clearly deviate. 

Interestingly, one of the YSR states ($\epsilon_4$) shows a particle-hole reversal in the spatial pattern of its \YSR excitations: the calculated  $|u_{\mathrm{n}} (\vect{r})|^2$  component matches the experimental map at negative bias. This reversal is an indication of this channel undergoing a  transition to a new correlated ground state  (as described in Fig.~\ref{fig:diagram}). As first pointed out by Sakurai~\cite{Sakurai70}, a critical point occurs for sufficiently large values of $J_\mathrm{n}\rho_s$, such that  the YSR  excitation energy becomes zero. This results in a new ground state of the many-body state (Fig.~\ref{fig:diagram}). In this situation, the magnetic impurity becomes locally screened, as pictorially described in Fig.~\ref{fig:GS}. Indeed, for the Cr/Pb(111) system, DFT simulations~\cite{Choi16} showed that the alignment of the \YSR excitations in the gap is very sensitive to the degree of hybridization $\Gamma_\mathrm{n}$, the Coulomb constant $U_\mathrm{n}$, and the energy of the spin-polarized state. It is thus possible that a fraction of scattering channels undergoes the transition to a screened singlet state, while others remain in the doublet ground state.

\section{Quantum-phase transition from a Kondo-screened to a free-spin ground state}
\label{Sec:QP}
 
In Section~\ref{Sec:orbital} we reported how the reduced symmetry around the magnetic atom results in a lifting of the orbital degeneracy and an orbital-dependent exchange potential $J_\mathrm{n}$, which bears multiple bound states in the gap. In realistic systems,  the exchange scattering potential $J_\mathrm{n}$ also leads to  quantum fluctuations of the impurity spin due to scattering with conduction band electrons, bearing the Kondo effect. For a normal metal, Kondo scattering results in screening of the impurity magnetic moment and the formation of a singlet ground state. 
As discussed in Section~\ref{sec:theory}, the formation of the superconducting singlet state competes with the Kondo singlet formation. 
The result is the coexistence of Kondo screening with the formation of \YSR states. Changes in the relative strength of the superconducting pairing and the exchange scattering can drive the system through a quantum phase transition separating the two different magnetic ground states.

Ideal systems to study the increasing role of Kondo screening in a superconductor with exchange scattering are hybrid superconductor-quantum dot  three-terminal devices. Here, the \YSR physics is represented by Andreev bound states (this term is usually used in the quantum-dot community) at the superconductor-quantum dot interface and the strength of their coupling to the superconductor can be continuously tuned by external gating fields. 

\subsection{Multiple magnetic ground states in a magnetic molecular system}

\begin{figure}[t]
	\includegraphics[width=1\columnwidth]{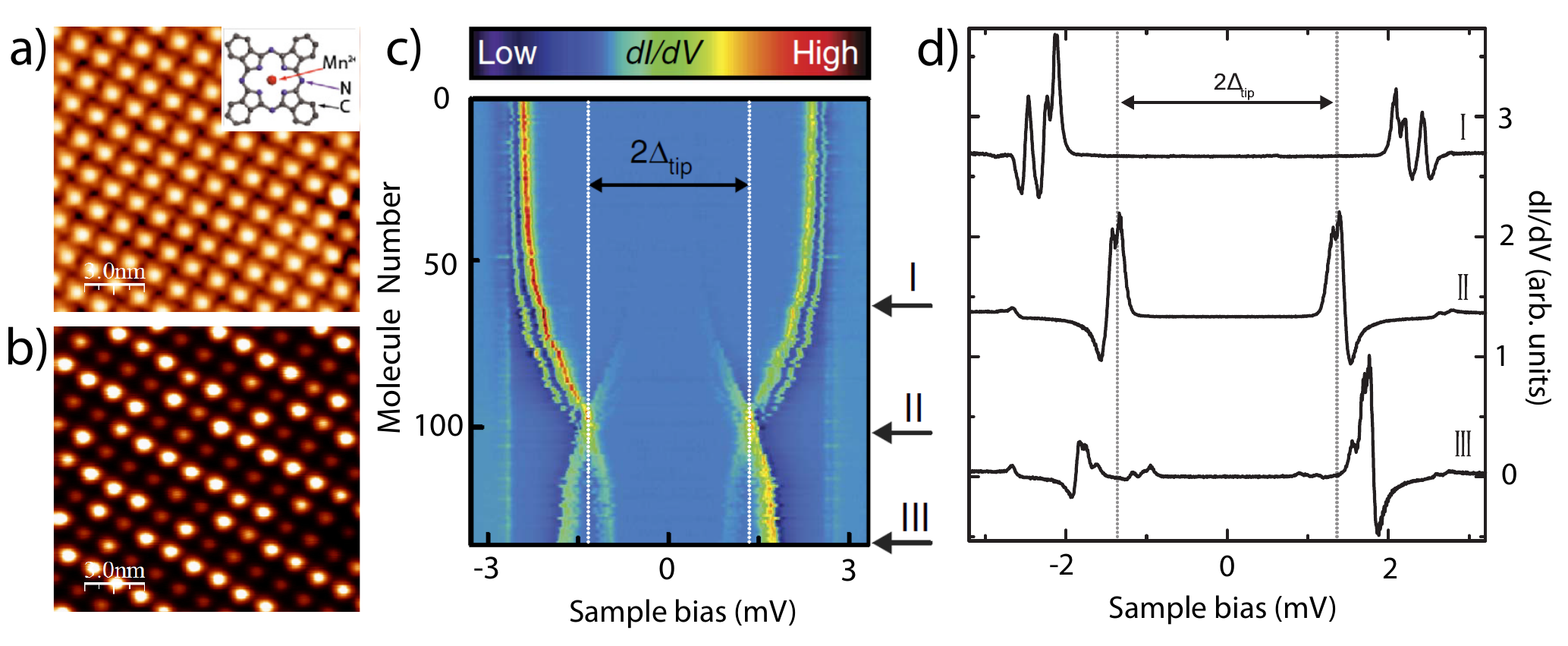}
	\caption{MnPc on Pb(111): (a) Constant current STM image and (b) the corresponding constant-height current map inside the superconducting gap ($V=1.2$\,meV) of an island of MnPc on Pb(111). The current map shows that each molecule has  a different "brightness" ({\it i.e.}, transmission)  depending on the alignment  of \YSR states with respect to the  coherence peaks of the superconducting tip. (c)  Color plot of stacked $dI/dV$ spectra of 137 MnPc molecules. The spectra are ordered from top to bottom following the energy position  of the most intense \YSR peak. The two ground states are indicated with arrows, as region I (Kondo ground state) and region III (free spin ground state). Region II corresponds to the crossing point between the two ground states. (d)  Three selected \didv spectra extracted from the stacked plot, one at each region of (c). Both, particle- and hole-like  \YSR excitation appear split into three peaks. As discussed in section 5.3, this multiplet is due to the effect of molecular anisotropy in the YSR states. Figure adapted from Ref.~\cite{Franke2011} (panel a and b)   and Ref.~\cite{hatter2015}  (panels c and d).}
	\label{fig:MnPc2}
\end{figure}

Another approach to trace the quantum phase transition was realised by taking advantage of a multitude of different adsoprtion sites of magnetic metal-organic molecules on top of a superconducting surface. Similar to atomic impurities, magnetic molecules such as transition metal phthalocyanines (Pc) on superconductor surfaces also show YSR excitations \cite{Franke2011,hatter2015,Kezilebieke2017}. The magnetism originates from  the incomplete $d$ shell of the metal ion and its exchange coupling with substrate electrons and cooper pairs. The  (organic) ligand field around the metal ion causes a finite spin anisotropy in these systems, favouring a certain spin orientation over others, and, hence, yields a characteristic excitation multiplet.  

A transition between two different ground states was observed by studying excitation spectra of MnPc molecules on Pb(111). MnPc molecules arrange in square molecular islands (Fig. \ref{fig:MnPc2}a), where each molecule lies over a  distinct atomic site, showing \YSR peaks  at different energy positions depending on this site.  The differences have been ascribed to small variations in the adsorption site, which crucially  affect the  strength of the exchange scattering.  The result is a  peculiar  Moir\'e pattern of interaction strength, which can be \textit{visualized} in   constant-height current images  at tunneling bias voltages inside the superconducting gap (Fig.~\ref{fig:MnPc2}b).

The diversity of possible sites of  MnPc on Pb(111)  allows us to explore a continuous range of  exchange interaction strengths $J$   in a single experiment simply by selecting different positions in the layer.   Figure \ref{fig:MnPc2} shows   a stack of spectra of $137$ different molecules ordered according to the position of the largest YSR spectral peak. We find a broad range of peak alignments, simulating  an experiment with a tunable  exchange  interaction. If we assume that the largest peak corresponds to the $u(r)$ component, a larger negative bias voltage denotes a larger exchange constant $J$. In this case, the plot is ordered from stronger to weaker $J$, from top to bottom. 
At $\pm\Delta$, the stack plot also shows the crossings of the \YSR resonances corresponding to the $u$ component with the thermal replica of the $v$ component and vice versa. 
This crossing  corresponds to the critical point mentioned above, where \YSR excitations reach zero energy and separate two different quantum ground states (Fig.~\ref{fig:diagram}). Above the critical point (region I), the larger interaction strength leads to a screened ground state, {\it i.e.}, a singlet (see Fig.~\ref{fig:GS}), and the \YSR peaks denote a doublet (single-particle) excitation. Below, the ground state correspond to a free-like impurity spin (doublet).     

\subsection{Interplay of Kondo correlations with  Yu-Shiba-Rusinov excitations}

As mentioned above, Kondo correlations participate in the screening of an  impurity spin on a superconductor, and their relevance in the ground state of a quantum spin depends on the strength of the exchange scattering.   In the regime where the Kondo energy scale  $k_\mathrm{B} T_\mathrm{K}$ competes  with the pairing energy $\Delta$, a  crossing of  two different quantum ground states was predicted by Matsuura~\cite{matsuura77}.
To probe the fundamental relationship between the \YSR energy $\epsilon$ and $k_\mathrm{B} T_\mathrm{K}$, both energy scales have to be determined simultaneously.

\begin{figure}[t]
\includegraphics[width=1\columnwidth]{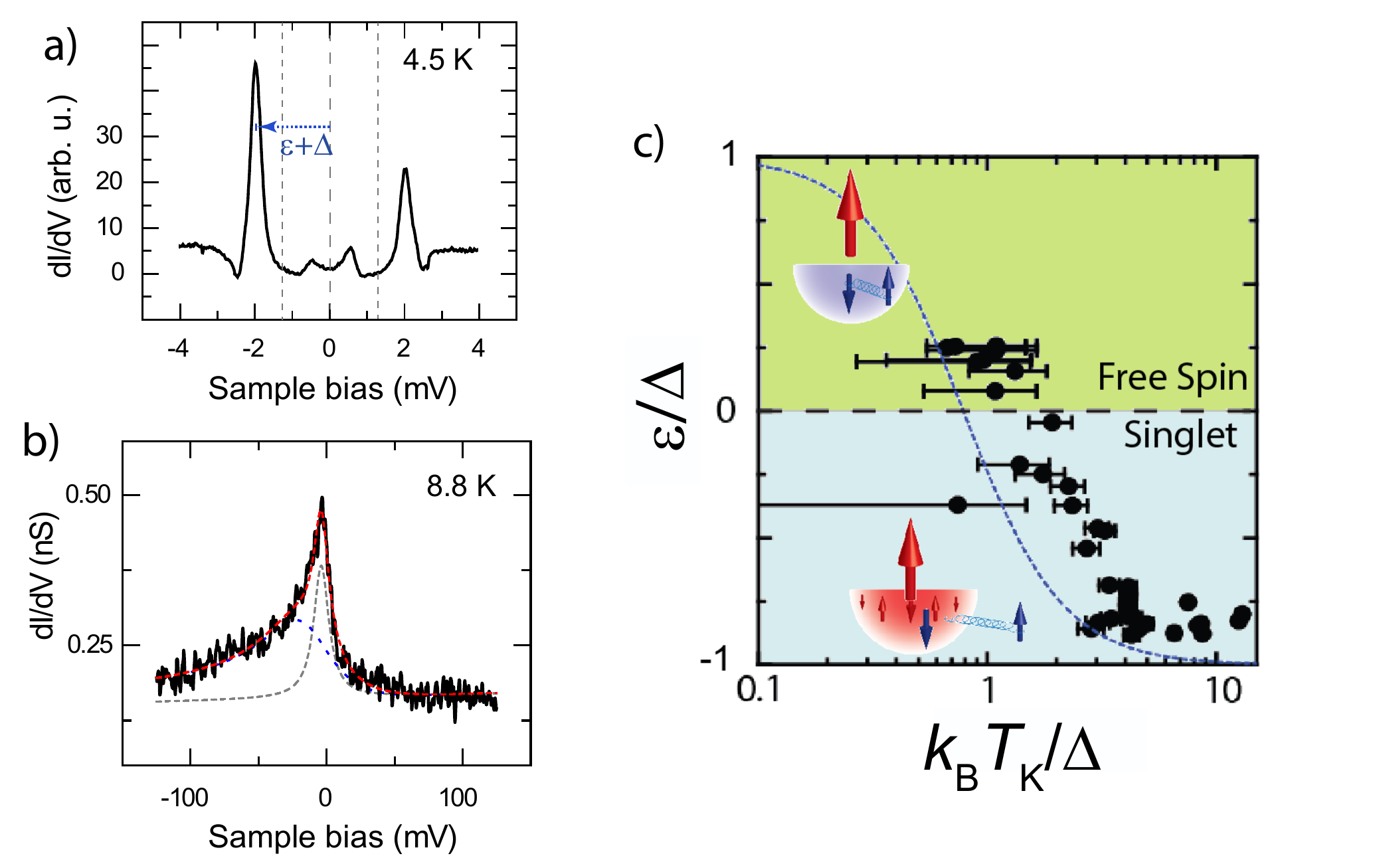}
    \caption{Relation between \YSR and Kondo energy scales for MnPc molecules: \didv spectra on a MnPc  molecule measured (a) at 4.5 K (below T$_c$) and (b) 8.8 K (above T$_c$), comparing their \YSR features with Kondo correlations. In (a), the hole-like \YSR excitation is larger, which signals for this system that it is in a singlet ground state.  In (b),  a zero-bias peak  allows us to quantify the strength of Kondo correlations. (c)  Correlation between the bound state energy ($\epsilon$) and the Kondo temperature (T$_k$) extracted from plots like in (a) and (b), respectively.  Dots are measured data points for  a set of MnPc molecules. The dashed line represents the predicted relation by Matsuura~\cite{matsuura77}. Figure adapted from Ref.~\cite{Franke2011}.   }
    \label{fig:phaseT}
\end{figure}  

For the case of MnPc,  $dI/dV$  plots measured above the critical temperature ($T_\mathrm{c}$) of Pb(111) show a characteristic zero-bias resonance attributed to the Kondo effect (Fig.~\ref{fig:phaseT}b). The  linewidth again  depends on the molecule investigated. A second (broader) resonance (fitted by a blue dashed line in the plot), which was originally interpreted as an additional Kondo channel, has been recently attributed to a Mn $d$ state \cite{Kugel2015}.   Comparison of spectra on each molecule above and below $T_\mathrm{c}$ (Fig.~\ref{fig:phaseT}a and  \ref{fig:phaseT}b) confirmed a correlation between the Kondo temperature and the \YSR excitation energy~\cite{Franke2011}. Figure~\ref{fig:phaseT}c plots the energy position 
$\varepsilon$ of the larger YSR peak vs. $k_\mathrm{B} T_\mathrm{K}$ as extracted from spectra like the ones shown in Fig.~\ref{fig:phaseT}b.  The \YSR peak shifts  towards more positive values as the Kondo energy scale becomes smaller. This corroborates that both spin-scattering processes depend similarly on the magnitude of exchange scattering $J$, and their relative strength approaches closely the  relation predicted by Matsuura \cite{matsuura77}. The crossing of the YSR peaks through zero occurs for $T_\mathrm{K} \sim \Delta$ and reveals the quantum phase transition. For stronger exchange, the ground state is a Kondo singlet. 
Beyond the critical point, Kondo correlations cannot screen the spin due to the depletion of states inside the superconducting gap ({\it i.e.}, $k_B T_K<\Delta$). In this case, the competing pairing correlations in the superconductor dominate and the  impurity's ground state transforms into a free-spin (doublet).

The exchange coupling to the substrate can be further tuned by the reversible addition of an axial ligand. An ammonia (NH$_3$) molecule bonding to the Mn ion increases the Mn--surface distance and reduces the exchange coupling $J$ such that all MnPc-NH$_3$ complexes are in the free-spin ground state. The concomitant Kondo effect then is in the weak-coupling regime~\cite{Hatter2017}.

\begin{figure*}[t]
	\includegraphics[width=1.5\columnwidth]{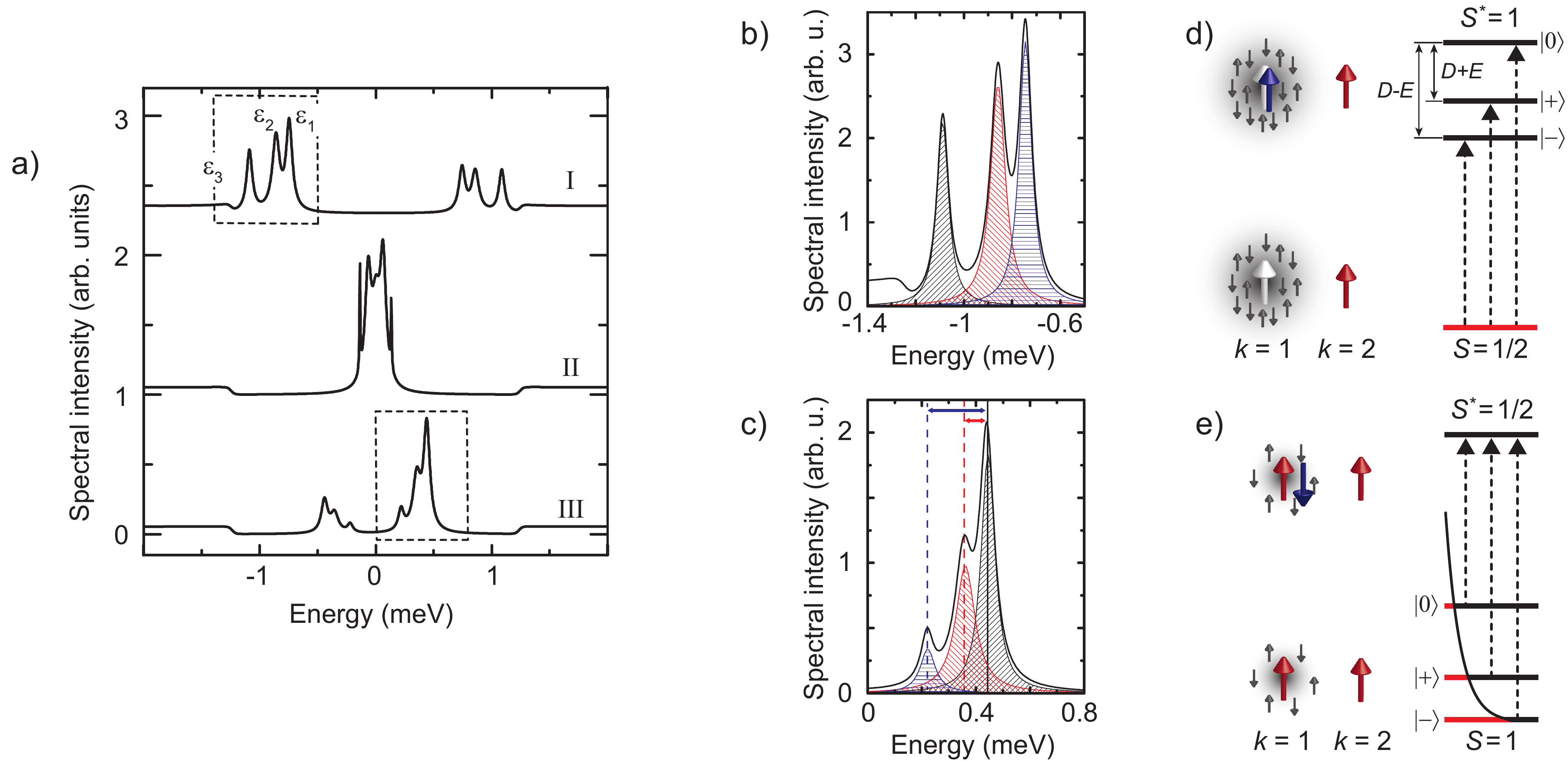}
	\caption{ (a)   Spectral intensity obtained by deconvolution of the three selected spectra from Fig. \ref{fig:MnPc2}d.  
The bound state resonances are labelled $\epsilon_1$,  $\epsilon_2$ and  $\epsilon_3$, respectively. Dashed boxes highlight the parts in the spectra relevant for the quantitative analysis of the peak areas and splittings. 
(b) and (c) shows a zoom  [marked in (a)] on three peaks for regions I (Kondo screened ground state) and III (free-spin ground state), with the corresponding components extracted as Lorentzian peaks. The area relation between the three peaks in the  later case follows close to a Boltzmann-like distribution for the temperature close to the experimental one.  (d) and (e) Illustration of the many-body states and the related energy
			level diagrams. (d) In the Kondo-screened ground state, the spin (white arrow) in scattering channel $k=1$ (the dominating channel) is screened, illustrated by grey arrows and
			shading, and the tunneling electron can enter with its spin (blue arrow)
			parallel to the spin in $k=2$ (the "hidden" scattering channel), increasing the excited state's total spin to $S^*$=1.
			The excitation scheme including the anisotropy-split excited state is shown
			by the energy-level diagram with indicated anisotropy parameters D and E.
			(e) In the free-spin ground state, the spin in $k=1$ (red arrow) is only
			partially screened. The tunneling electron enters this state in an
			anti-parallel alignment, obeying the Pauli exclusion principle and decreasing
			the spin in the excited state to $S^*$=1/2. Figure adapted from Ref.~\cite{hatter2015}. }
	\label{fig:ani}
\end{figure*}

\subsection{Effect of magnetic anisotropy in Yu-Shiba-Rusinov states}
\label{SubSec:anisotropy}

The spectra in Fig. \ref{fig:MnPc2}d  show that \YSR resonances in MnPc   appear split into three narrow peaks, typically with $50-100$\,$\mu$eV full width at half maximum,  with different intensity, and separated by up to 400\,$\mu$eV.   The  split peaks are  a result of the intrinsic magnetic anisotropy of the MnPc molecule, as predicted by 
  \v{Z}itko and coworkers \cite{Zitko2011}, 
and demonstrated in Ref. \cite{hatter2015}. When the magnetic impurity has a spin larger than 1/2, intra-impurity  magnetic anisotropy can split  its energy levels into a  multiplet of $S_z$ components. The spin 3/2 of free MnPc molecules is partially decreased to $S_{imp}=1$ at the Pb(111) surface  \cite{Jacob2013a}, with two spin-polarized orbitals, $d_{z^2}$ and $d_{xy}$. The exchange interaction with the surface  is dominated by the $d_{z^2}$ orbital, whereas the $d_{xy}$ state remains "hidden" in the molecular plane. Even though there is only one tunneling and scattering channel (the $d_{z^2}$ orbital),  correlations between the two spin-polarized orbitals split the corresponding \YSR state  into a multiplet. The energy split of the multiplet reflects the magnetic anisotropy of the impurity spin, but is renormalized by the many-body interactions.

However, the essence behind the fine structure is more complex \cite{hatter2015}. As pictured in Fig. \ref{fig:transport} in Sec.~\ref{Sec:transport}, transport in the regime of single particle tunneling involves the formation of an excited many-body state, the \YSR excitation, with a different occupation than the ground state. 
The lifetime of the excited state is, as shown in Sec.~\ref{Sec:transport}, in the order of hundreds of ps at 1.2\,K. This reduces the excitation linewidth such that the measurement becomes sensitive to the magnetic anisotropy of the excited state. 
Hence, the anisotropy splits the excited state into different energy levels.

The spectral multiplet thus shall reflect all possible transitions connecting the many-body ground state multiplet with the excited state multiplet, ensuring conservation of the total angular momentum including the spin of the tunneling electron. We note that, while the population of the many-body ground state's  multiplet should follow a Boltzmann statistics, the excited state's multiplet can be accessed with similar probability. Therefore, the amplitudes of the peaks in the Shiba substructure shall reflect the strength of each possible transition, allowing to determine its origin. 

The   sequence of spectral peaks in the deconvolved spectra depicted  Fig. \ref{fig:ani}a  illustrates the peculiar variation of the spectral intensity of each peak according to the ground state.  As outlined in  Figs.  \ref{fig:ani}b and \ref{fig:ani}c, in the Kondo screened case (region I in  Fig. \ref{fig:MnPc2}c) all spectra show peaks with similar intensity, independently of their position. On the contrary, the spectral intensity of the  peaks in the free-spin ground state (region III)  follows a thermal-like distribution of intensities: hole and particle  excitations become stronger with their energy, following  a Boltzmann-like distribution of the peak areas \cite{hatter2015}. From the arguments above, this signals that they correspond to transitions from a multiplet in the $S=1$ ground state. This triplet state originates from the intra-atomic magnetic exchange of the scattering channel ($k=1$ in Fig. \ref{fig:ani}e) and the hidden spin ($k=2$), while the single-particle tunneling via the  $k=1$ channel leads to a $S^*$=1/2 excited state. 

On the contrary, the similar intensity of triplet excitations in the Kondo-correlated ground state is a fingerprint of transitions to an excited-state multiplet  with an excited spin $S^*$=1. As depicted in Fig.  \ref{fig:ani}d, this is fully consistent  with the existence of the "hidden" spin in the $k=2$ channel accompanying the screened channel $k=1$. It is remarkable that, due to the long lifetime of the excitations, the \YSR excited state is sensitive to the intrinsic magnetic anisotropy of the molecule and split into three levels. It is worth mentioning that, since these results probe  properties of the \YSR many-body state, the observed split corresponds to the intra-molecular anisotropy but renormalized by the exchange scattering interactions.

\section{Lateral extension of Yu-Shiba-Rusinov states}
\label{Sec:extension}

\begin{figure*}[t]
\includegraphics[width=1.5\columnwidth]{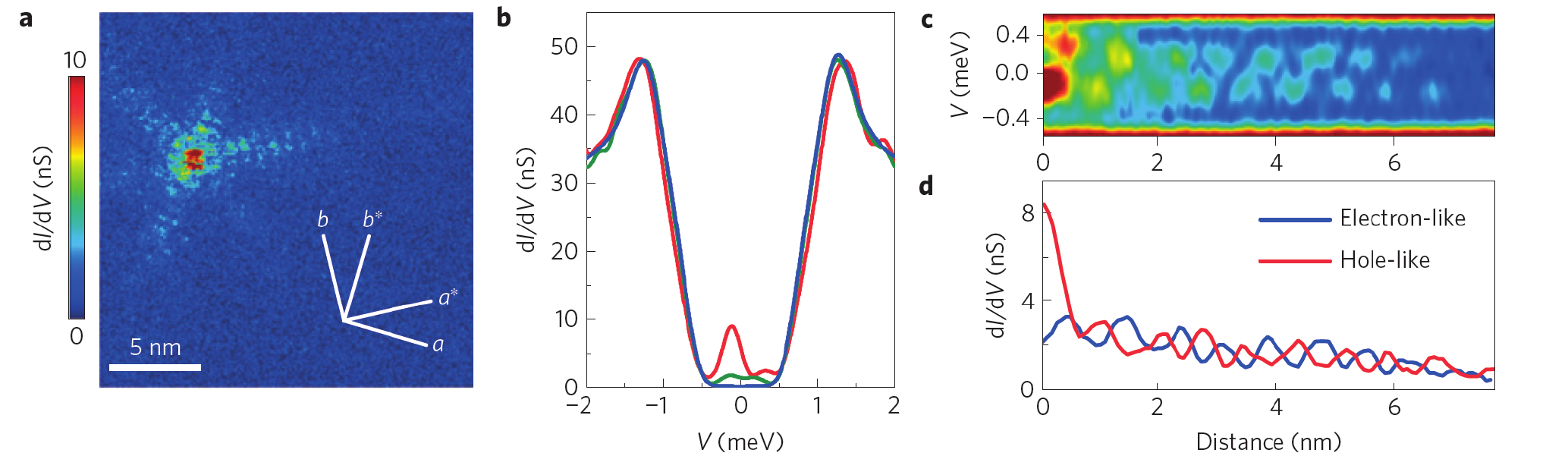}
\caption{Subsurface impurities in 2H-NbSe$_2$. a) $dI/dV$ map of subsurface impurities in 2H-NbSe$_2$ show an oscillatory falloff of the S\YSR wave function. b) $dI/dV$ spectrum above the center of the scattering pattern. c) pseudo-2D representation of $dI/dV$ spectra as a function of distance. d) Oscillatory falloff of the electron- and hole-component of the Shiba wave function. Reprinted by permission from Macmillan Publishers Ltd: Nature Physics~\cite{Menard2015}, copyright 2015.}
\label{fig:Menard}
\end{figure*}

In section \ref{Sec:orbital}, we discussed the shape of the \YSR states in the close vicinity of the impurity, which reflects the orbital shape of the scattering potential. Moreover, the \YSR states show a longer-ranged intensity, which is determined by several characteristic length scales of the superconductor: On the one hand, the coherence length $\xi$ plays a role, as expected intuitively for a superconducting material. The decay of the \YSR wave function scales as $\exp(-r/\xi)$ (see Eq.~\ref{eq:ShibaWF}) with the distance $r$ from the impurity. On the other hand, scattering at an impurity is governed by fermionic scattering processes, which occur on the length scale of the Fermi wavelength. The physical concept relates to the Friedel-like screening of the impurity site. 
The \YSR wave function thus exhibits an oscillation with the Fermi wavelength $\lambda_\mathrm{F}$ and an additional $1/(k_\mathrm{F}r)$ decay for an isotropic 3D superconductor (see Eq.~\ref{eq:ShibaWF}). We first discuss the contributions to the decay. Typical type I superconductors possess a coherence length in the order of 100~nm, whereas the Fermi wavelength is only in the order of 1~nm. Hence, the decay at close distance around the impurity will be dominated by the Fermi wavelength and, in particular, by the dimensionality of the superconducting substrate, whereas the coherence length has minor influence. Menard and co-workers showed on the quasi-2D superconductor NbSe$_2$ that one can indeed observe a long-range extension of \YSR states \cite{Menard2015}. Sub-surface impurities led to an oscillation pattern of the \YSR states with significant amplitude of the wave function up to 7~nm  away from the impurity site (see Fig.~\ref{fig:Menard}). This was explained by the reduced damping of the wave function in 2D compared to 3D, with $u(r)$, $v(r)$ scaling like $1/\sqrt{k_\mathrm{F}r}$ in 2D. 

\begin{figure}[b]
\includegraphics[width=1\columnwidth]{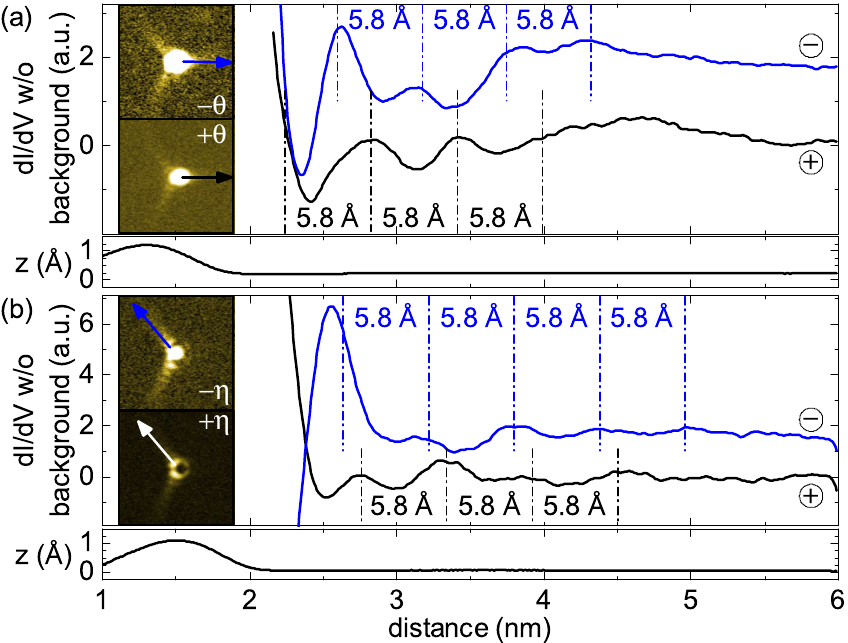}
\caption{Mn adatoms on Pb(111). Oscillatory modulation of the \YSR wave function for two pairs of bound states [$\pm\theta$ (a) and $\pm\eta$ (b)]. Figure adapted from Ref.~\cite{ruby2016}.}
\label{fig:oscillation}
\end{figure}

\subsection{Enhancement of \YSR oscillations by an anisotropic Fermi surface}
In contrast, Pb is a 3D superconductor. Hence, one may expect a much faster decay of the wave function. Surprisingly, Mn atoms on Pb(111) also showed patterns of \YSR states up to 4 nm away from the adsorption site (see Fig.~\ref{fig:oscillation}). It is, however, interesting to note that the \YSR states of Mn on Pb(100) (see Fig.~\ref{fig:orbital}) were much shorter ranged (only up to 2 nm). Hence, the surface orientation seems to play an important role. Furthermore, we note that the decay is not spherically symmetric, but "beams" extend along the high symmetry directions of the surface. This holds true for 2D and 3D superconductors and  can be ascribed to the anisotropy of the band structure of the substrate \cite{Salkola1997,Flatte97PRB}.  Electrons/holes scattered at an impurity are focused into directions, which originate from flat areas of the Fermi surface \cite{Weismann2009}. This focusing effect enhances the \YSR amplitude for these directions and gives rise to a fall-off, which is slower than expected for an isotropic decay.  Hence, we can observe the \YSR wave function nanometers away from the impurity. 

The fermionic properties of the scattering processes are also imprinted in an oscillatory pattern of the \YSR states. Hence, the wave function does not only decay with the characteristic $1/(k_\mathrm{F}r)$ dependence, but also oscillates with the Fermi wavelength $\lambda_\mathrm{F}$. Because STM probes the probability density $\left|\Psi\right|^2$, the observed oscillation is periodic in $\lambda_\mathrm{F}/2$. Fig.~\ref{fig:oscillation} reveals an oscillation around the Mn adatom on Pb(111) with a period of $\sim 5.8$~\AA, which thus reflects $\lambda_\mathrm{F}\sim 11.6$~\AA. Pb exhibits a complex Fermi surface with two disjoint sheets with significantly different values of the Fermi wave vector $k_\mathrm{F}$ along the different lattice directions \cite{Lykken1971}. Importantly, the electron-phonon coupling strength is different on the two sheets and gives rise to two distinct superconducting energy gaps \cite{ruby2015a,Floris2007}. The observation of a single, well-defined  oscillation period evidences that the magnetic impurity is predominantly coupled to the corresponding Fermi surface sheet. In this case of Mn on Pb(111), it reveals the coupling of the impurity spins to the Fermi surface sheet with mainly $p-d$ orbital character instead of to the Fermi surface sheet with mostly $s-p$ orbital character. Considering the more local nature of the electrons in the $p-d$-like bands, they are  affected more strongly by a local impurity than extended bands. More importantly, in Sec.~\ref{Sec:orbital} we have shown that dominantly $l=2$ fermions scatter with the unpaired spins in the $d$ states of Mn as was early predicted by Schrieffer in the case of Kondo scattering~\cite{Schrieffer1967}. 
Hence, the restriction to certain angular momenta for efficient scattering favors the $p-d$-like band.
The scattering pattern then is a result of the interplay of orbital symmetry and Fermi surface sheet. While at close distance, the orbital symmetry dominates the pattern, at far (and intermediate) distances, the electron focusing from the flat parts of the Fermi surface prevails.  

Interestingly, the scattering pattern of the \YSR states obey distinct phase differences between the $u$ and $v$ components. The observed phase difference~\cite{ruby2016} depends on the energy of the \YSR state and follows the theoretical prediction  of Eq.~\ref{eq:shift} of a point scatterer. 

It is noteworthy that, in the case of Cr atoms on Pb(111) films grown on SiC(0001)~\cite{Choi16}, no oscillatory fall-off was observed. Concurrently, no signs of two-band superconductivity were observed for the Pb films, in contrast to the single-crystal samples~\cite{ruby2015a}. Both findings could be linked to the insufficient film thickness, which does not allow for the $p-d$-like band to be fully developed. Hence, there is no preferential scattering with a single band and the focusing effect is absent. Hence, the \YSR wave function decays much faster in this case.
\\
\section{Coupling of Shiba impurities -- from dimers to topological chains} 
\label{Sec:coupling}

The extension of the \YSR states is a promising avenue for coupling magnetic impurities on a superconductor. 
Theoretically, the coupling of \YSR states has already been discussed since the advent of the understanding of the individual states \cite{Rusinov1968}. The physical picture is based on the spatial overlap of \YSR states. The \YSR states thus hybridize, with the hybrid state being described by the linear combination of the \YSR states (similar to LCAO theory or to the description of interference patterns due to the overlap of oscillating \YSR states \cite{Morr2003, Morr2006}). Experimentally, hybrid \YSR states have first been seen on dimers of Mn and Cr on Pb(111) \cite{Ji2008}. Tunneling spectra revealed a splitting of the \YSR states. The corresponding STM images reflected the bonding (anti-bonding) nature of the coupled states via a maximum (zero) intensity at midway between the impurity sites. Recently, a splitting of \YSR states has also been observed on dimers of Co-Phthalocyanine molecules adsorbed on the quasi-2D superconductor NbSe$_2$ \cite{Kezilebieke2017} and on dimers of Cr adatoms on the surface of the superconductor $\beta$-Bi$_2$Pd \cite{ChoiArxiv2017}. 

\begin{figure}[t]
\includegraphics[width=1\columnwidth]{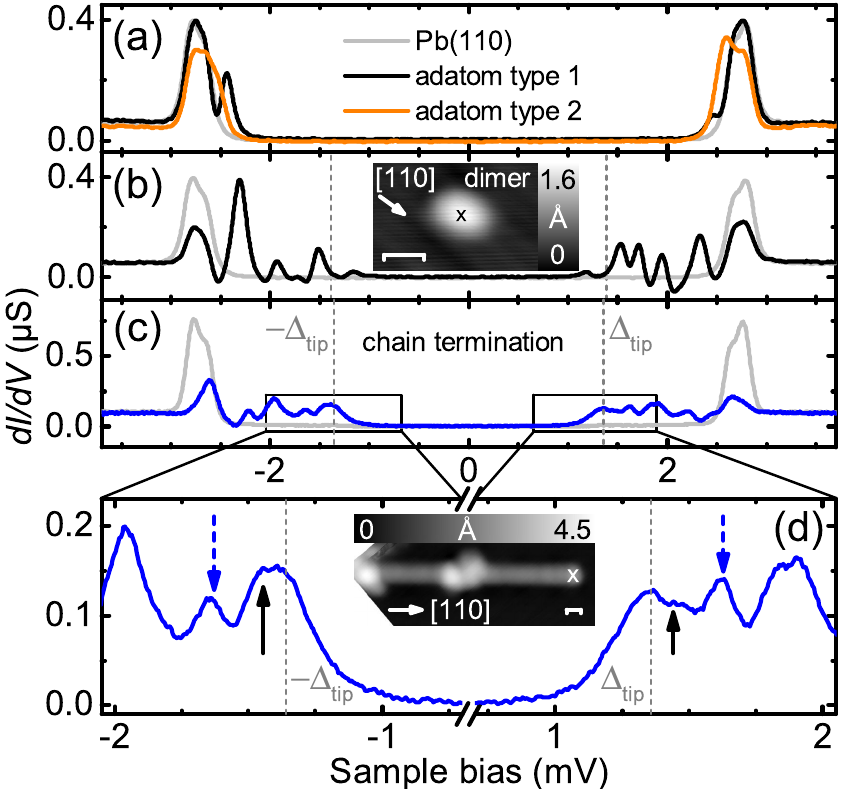}
\caption{Fe on Pb(110). $dI/dV$ spectra acquired above two adatom species (a), above an Fe dimer (b), and above the end of an iron chain (c and d). Figure adapted from Ref.~\cite{Ruby2015b}.}
\label{fig:Fe}
\end{figure}

\subsection{Chains of Yu-Shiba-Rusinov impurities}
In the limit of dense arrangements of magnetic impurities, the \YSR states' overlap may lead to extended bands. These fill and eventually suppress the superconducting energy gap \cite{Ji2010a}. An intriguing example for the formation of \YSR bands are one-dimensional chains of transition metal adatoms on a superconducting Pb surface. Employing a Pb(110) surface steers the assembly of Fe~\cite{Nadj-Perge2014,Ruby2015b,Pawlak2016,Feldman2016} and Co~\cite{Ruby2017} chains along the troughs of the surface. Whereas dimers of  Fe atoms in these troughs already show an increase in the number of \YSR resonances in the superconducting energy gap (Fig.~\ref{fig:Fe}b), the chains show a rich resonance structure (Fig.~\ref{fig:Fe}c). While the dominant resonances can be interpreted as van Hove singularities of \YSR bands, the rich and varying structure along the chain results from confinement effects in the finite chain and variations in the local potential~\cite{Ruby2015b,Ruby2017,Peng2015}. 

\begin{figure*}[t]
\includegraphics[width=1.5\columnwidth]{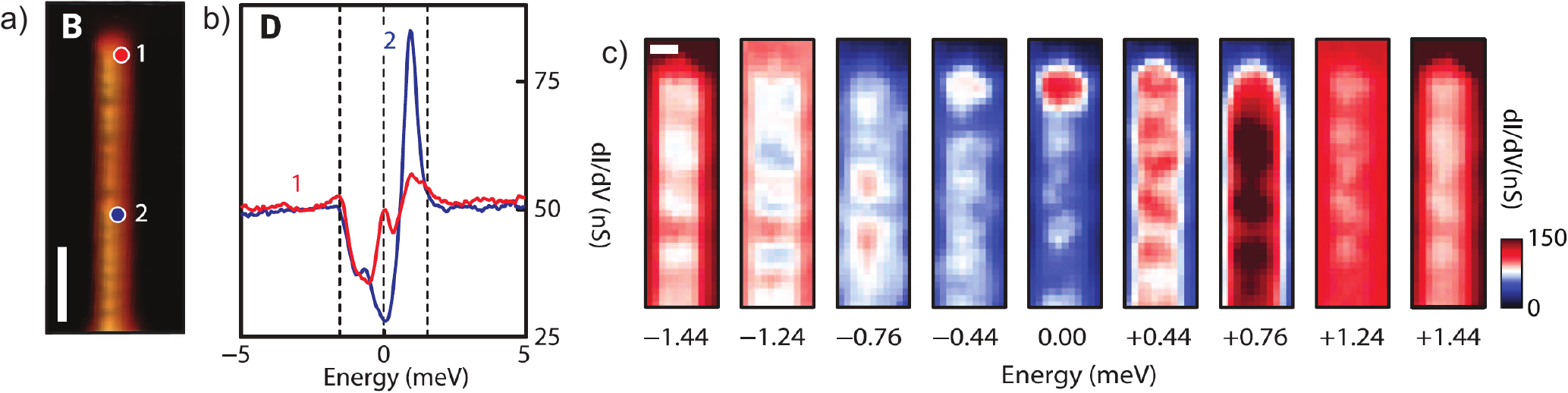}
\caption{Fe chains on Pb(110). b) $dI/dV$ spectrum above the center (2)  and the end (1) of the iron chain shown in (a).  c) $dI/dV$ maps of the iron chain at different energies around $E_F$. The zero energy map shows an intensity localization at the chain end. From Ref.~\cite{Nadj-Perge2014}. Reprint with permission from AAAS.}
\label{fig:Nadj}
\end{figure*}

Besides the \YSR bands at finite energy, which are found all along the chain, Nadj-Perge and co-workers identified a resonance at zero energy, which is localized at both chain terminations (Fig.~\ref{fig:Nadj}). These resonances have been interpreted as Majorana zero modes~\cite{Nadj-Perge2014}.

Topological superconductivity is a crucial prerequisite for the emergence of Majorana zero modes. On an $s$-wave superconductor, this can be realized either by helical spin chains ~\cite{Nadj-Perge2013,Pientka2013,Braunecker2013, Klinovaja2013,Vazifeh2013,schecter2016,Christensen2016} or by ferromagnetic chains in the presence of strong spin-orbit coupling, as it occurs in Pb~\cite{Nadj-Perge2014, Li2014, Peng2015}. The induced $p$-wave superconducting gap protects the topological nature of the band structure. The first putative realization of such a system are the above mentioned Fe chains on Pb(110), where an odd number of spin-polarized band cross the Fermi level. These provide the origin of the  Majorana states at the chain ends \cite{Nadj-Perge2014,Li2014}. 

These zero energy modes are the condensed matter equivalent to Majorana Fermions~\cite{Majorana1937} in particle physics. They possess numerous interesting properties, in particular, an non-local character (in the case of the chains, they always appear at both terminations simultaneously), topological protection against perturbations, and obeying non-Abelian exchange statistics~\cite{Alicea2012,Beenakker2013,Elliott14}. These properties make them prime candidates for quantum computational applications~\cite{Sarma2015}. In the simplest approach, Majorana zero modes could be used to store quantum information non-locally. More evolved proposals use networks of Majorana wires in order to allow for braiding operations to realize certain quantum gates~\cite{Alicea2011}.

The prospect of using Majorana states for topological quantum computing has driven the investigation of the formation of Majorana bound states in proximity-coupled chains of transition metal atoms on a superconductor to the forefront of research in solid state physics. 
Yet, braiding operations, \ie, the exchange of Majorana modes in, say, a Y-junction requires the physical movement of the zero energy mode along the wire. In the case of atomic wires on an $s$-wave superconductor, both for ferromagnetic and chiral order, the topological phase is determined by the chemical potential relative to the energy of the spin-split bands. A phase transition between a trivial and a nontrivial phase occurs when the chemical potential moves through the bottom of a spin-polarized band, because this changes the number of Fermi points. Hence, in order to move the Majorana mode located at the domain wall between a topologically trivial and a nontrivial phase, one has to tune the chemical potential on the energy scale of the $d$ band width, \ie, some hundreds of meV~\cite{Oppen2017}. Technologically, such braiding operations would be more than challenging. A loophole to a successful applications could lie in the  creation of ring-like quasi-one-dimensional structures on a thin-film superconductor. If the topological phase is then controlled by the amplitude and orientation of a (dynamic) magnetic field, an in-plane rotation of the field can move the topological domain wall and hence the Majorana zero mode, making braiding experiments possible~\cite{Li2016}.
\\

\section{Conclusions and future directions}
\label{sum}
In summary, we described the latest experimental results on individual magnetic adsorbates on $s$-wave superconductors. In combination with theory, tremendous advances have been made in the fundamental understanding of the origin of multiple subgap excitations and their relation to the many-body quantum states. The investigations also bridged the gap between STM-based research and quantum nanoelectronics. Despite of this progress, new experiments are required to go beyond the current state-of-the-art, in particular in view of the construction of adatom-based nanostructures. One interesting aspect is the coupling between adatoms, which is currently treated by theory in great detail. The coupling mechanism, its length scale, as well as resulting magnetic ground states are modeled~\cite{Morr2003, Meng2015, Hoffman15}. Yet, experimental evidence of the different scenarios is mostly still missing. While these questions are again of fundamental nature, they also form the basis for understanding the coupling in larger adatom structures, eventually leading to one-dimensional and two-dimensional topological structures.

\section*{Acknowledgments}
We are thankful for fruitful discussion with Gelavizh Ahmadi, Tristan Cren, La\"etitia Farinacci, Nino Hatter,  Nicolas Lorente, Felix von Oppen, Yang Peng, Falko Pientka,  Ga\"el Reecht, Dimitri Roditchev, Michael Ruby, Gunnar Schulze, and Clemens Winkelmann. We acknowledge financial support by the Deutsche Forschungsgemeinschaft by Grant No. FR2726/4 (K. J. F.) and Grant No. HE7368/2 (B. W. H.), by the European Research Council through Consolidator  Grant {\it NanoSpin} (K. J. F.).

\bibliographystyle{apsrev4-1}
\bibliography{biblio_new}{}

\end{document}